\documentclass[12pt]{iopart}
\usepackage{iopams}
\usepackage{theorem}
\usepackage[dvips]{graphicx}
\usepackage{amssymb}
\usepackage{euscript}

\newcommand{\B}{\mathfrak{B}}

\newcommand{\Hi}{\mathcal{H}}

\newcommand{\rank}{\mathop{\rm rank}}

\newcommand{\lv}{\left \vert} 
\newcommand{\rv}{\right \vert}
\newcommand{\la}{\left \langle}
\newcommand{\ra}{\right \rangle}
\newcommand{\ket}[1]{\lv #1 \ra}
\newcommand{\bra}[1]{\la #1 \rv}
\newcommand{\braket}[2]{\la #1 \vert #2 \ra}

\theorembodyfont{\rmfamily}
\newtheorem{Theorem}{{Theorem} }
\newtheorem{Remark}{Remark}

\newtheorem{Corollary}{{Corollary}}
\newtheorem{Lemma}{{Lemma}}
\newtheorem{Proof}{{Proof}}

\begin{document}

\title{Two-way classical communication remarkably improves local distinguishability}

\author{Masaki Owari $^{1,2}$ and Masahito Hayashi $^{3,4}$}

\address{$^1$ Collaborative Institute of Nano Quantum Information Electronics, The University of Tokyo  \\
Hongo 7-3-1, Bunkyo-ku Tokyo 113-0033, Japan\\
$^2$ Department of Physics, Graduate School of Science, The University of Tokyo \\
Hongo 7-3-1, Bunkyo-ku Tokyo 113-0033, Japan\\
$^3$ ERATO-SORST Quantum computation and information project, JST,
Hongo 5-28-3, Bunkyo-ku Tokyo 113-0033, Japan\\
$^4$ Graduate School of Information Sciences, Tohoku University, Aoba-ku, Sendai, 980-8579, Japan}

\ead{\mailto{masakiowari@is.s.u-tokyo.ac.jp}, \mailto{hayashi@math.is.tohoku.ac.jp}}
\begin{abstract}
We analyze the difference in the local distinguishability among the following three restrictions; (i) Local operations and only one-way classical communications (one-way LOCC) are permitted. (ii) Local operations and two-way classical communications (two-way LOCC) are permitted. (iii) All separable operations are permitted. We obtain two main results concerning the discrimination between a given bipartite pure state and the completely mixed state with the condition that the given state should be detected perfectly. 
As the first result, we derive the optimal discrimination protocol for a bipartite pure state in the cases (i) and (iii). As the second result, by constructing a concrete two-way local discrimination protocol, it is proven that the case (ii) is much better than the case (i), i.e., two-way classical communication remarkably improves the local distinguishability in comparison with one-way classical communication at least for a low-dimensional bipartite pure state.
\end{abstract}

\pacs{03.65.Ud,03.65.Wj,03.67.Mn}
%03.65.Ud Entanglement and quantum nonlocality  
%03.65.Wj State reconstruction, quantum tomography  
%03.67.Mn Entanglement measures, witnesses, and other characterizations  
%02.50.Le Decision theory and game theory  
%02.50.Tt Inference methods 
%03.67.-a Quantum information  
%03.67.Bg Entanglement production and manipulation  

\vspace{2pc}
\noindent{\it Keywords}: 
state discrimination,
one-way LOCC,
two-way LOCC,
Schmidt rank,
Schmidt coefficient
% Uncomment for Submitted to journal title message

\submitto{\NJP}
\maketitle
\section{Introduction}
Recently, quantum communication has been 
investigated among many groups as a future technology.
Similar to conventional information technology,
practical quantum communication technology will require
distributed information processing among two or more spatially separated parties.
In order to treat this problem, 
it is necessary to clarify what kind of information processing is possible
under respective constraints for permitted operations.
In the quantum case, 
when our quantum system consists of distinct two parties $A$ and $B$,
we often restrict our operations to 
local (quantum) operations and classical communications (LOCC)
because sending quantum states over long distance is technologically more difficult than sending classical information\cite{basic paper of entanglement}.
Even in this restriction, we can consider the following two formulations;
(i) The classical communication is restricted to the direction $A$ to $B$
(We can similarly treat the restriction of the opposite direction.)
(ii) All parties are allowed to communicate classically with each other as much as they like. 
The case (i) is called the one-way LOCC, and the case (ii) is called the two-way LOCC. 

Since, by definition, the two-way LOCC apparently includes the one-way LOCC,
the two-way LOCC is always more powerful than the one-way LOCC in principle.
However, due to the following two reasons, it is not easy to characterize the difference between the both performances.
As the first reason, it is mathematically hard to rigorously evaluate the
performance of a given two-way LOCC protocol,
because the mathematical description of two-way LOCC is too complicated.
As the second reason, for several simple tasks, the performance of the two-way LOCC
was actually shown to be the same as the performance of the one-way LOCC. 
In fact, there are several settings that has no difference between one-way LOCC and two-way LOCC, e.g., LOCC convertibility of bipartite pure state \cite{LP01}, 
Stein's lemma bound in the simple asymptotic hypothesis testing of the $n$-tensor product of identical states\cite{Hayashi text}.

On the other hand, in several settings, the two-way LOCC is strictly powerful than the one-way LOCC.
For example, the distillable entanglement (the amount of
maximally entangled states which can be derived from a given state
by LOCC) with two-way classical communication is proven to be
greater than that with one-way classical communication\cite{BDSW paper}. 
Also, this type comparison also has been done by several papers\cite{LD,GV,cohen} 
only on the discrimination among orthogonal states. 
Although several researchers treated this problem,
they did not treated the discrimination among non-orthogonal states.
In this paper, we compare the both performances quantitatively on 
the ``{\it local discrimination}''  among states that are not necessarily orthogonal, whose purpose is discriminating given states by only LOCC with the {\it single} copy.
In fact, general discrimination problem is closely related to 
sending classical information via quantum channel\cite{coding}
and quantum algorithm \cite{Ambainis,HKK}.

%This is because when we encode classical information $\{1, \cdots, n \}$ onto a set of quantum states $\{\rho _i \}_{i=1}^n$, we need to distinguish the set of states $\{ \rho _i \}_{i=1}^n$. 
%That is, information-theoretically, the distinguishability means how much classical information can be derived from an unknown state. 
%Therefore, if the performance of local discrimination with two-way classical communication is better than the performance with one-way classical communication, we can conclude that the two-way classical communication enhances the power to derive a classical information from an unknown state compared to the one-way classical communication. 
%However, there is no example which proves the benefit of two-way classical communication in the local discrimination problem;
%that is, almost all known local discrimination protocols can be implemented by only 
%one-way LOCC \cite{LD, VSPM01,two-state-LD}.
%while the local discrimination has been discussed by many researchers\cite{LD, VSPM01, 
%asymptotic paper, local discrimination, two-state-LD}.

In order to quantify the difference of the two-way LOCC and the one-way LOCC,
in this paper, we concentrate on a simple setting:
the local discrimination of the first state $\rho$ on a bipartite system $\Hi$ from the second state $\tilde{\rho}$ under the condition 
where the first state $\rho$ should be detected perfectly. 
When the both states $\rho$ and $\tilde{\rho}$ are pure,
there is no difference between one-way LOCC and two-way LOCC
because any global discrimination protocol can be simulated by one-way LOCC\cite{VSPM01, two-state-LD}.
Surprisingly, as our result, we found that there usually exists non-negligible difference between two restrictions when the second $\tilde{\rho}$ is the completely mixed state $\rho_{\rm mix}:=I/\dim \Hi$.
At the first glance, this setting seems specific, however,
due to the following six reasons, it is closely related to several research topics.
First, this type analysis produces a bound of the number of perfectly locally distinguishable states.
Second, as is explained later, there is a relation between the performance of local distinguishability and amount of entanglement in the case of pure states.
Third, this kind of distinguishability is often treated in quantum complexity as Triviality of Coset State \cite{HKK,TCS}.
Fourth, when the second state $\tilde{\rho}$ is close to the completely mixed state $\rho_{\rm mix}$, we obtain a similar conclusion because the power of our test is continuous concerning the second state. 
Fifth, in the community of classical statistics,
the problem of discriminating the given two distributions is widely accepted as
the fundamental problem of hypothesis testing
because general hypothesis testing problem can be treated by using this type problem\cite{lehmann}.
Sixth, as was mentioned in the preceding papers \cite{coding}, 
hypothesis testing with two candidates states is closely related to 
quantum channel coding. 
Hence, it is suitable to treat this kind of local discrimination problem.

In order to analyze this problem in the respective settings,
we introduce the minimum error probabilities to detect the complete mixed state 
$\beta_{\rightarrow }(\rho )$, $\beta_{\leftrightarrow }(\rho)$, and $\beta_{\rm sep }(\rho )$ 
by the one-way LOCC, the two-way LOCC, and the separable operations, respectively. 
Indeed, these functions are considered as appropriate measures of the local distinguishability because they give not only the minimum error probability of the above problem, but also the upper bound of the size of locally distinguishable sets in general perfect local discrimination problems\cite{local discrimination}. 
Under this formulation, we first analyze the local distinguishability by means of one-way LOCC and separable operations, and derive the optimal discrimination protocol with one-way LOCC and separable operations;
we should note that the minimum error probability $\beta_{\rm sep }(\rho )$ 
with separable operations gives a lower bound for 
the minimum error probability $\beta_{\leftrightarrow }(\rho)$ with two-way LOCC. 
After that, constructing a concrete two-way local discrimination protocol, we show that two-way classical
communication remarkably improves the local distinguishability in
comparison with the local discrimination by one-way classical
communication at least for a low-(less than five) dimensional bipartite pure state.
Indeed, since the power of our test is continuous concerning the first and the second states,
our result indicates that 
two-way classical communication remarkably improves the local distinguishability 
in a wider class of the first and the second states.
Moreover, as a byproduct, we extend the characterization of 
locally distinguishability by one-way LOCC by Cohen \cite{cohen} to a set of mixed states.

This paper is organized as follows: In Section \ref{preliminary},
we introduce the discrimination problem between an arbitrary given state $\rho$ and a completely mixed state $\rho_{\rm mix}$ on a bipartite system $\Hi$ under 
the condition that the given state is detected perfectly.
Then, we explain another meaning of $\beta_{\rightarrow }(\rho )$, $\beta_{\leftrightarrow }(\rho )$, and $\beta_{\rm sep }(\rho )$ 
from the viewpoint of general local discrimination problems. 
In Section \ref{separable section chap.3}, constructing the optimal separable POVM for the local discrimination, we prove that $D \beta_{\rm sep }(\ket{\Psi} )-1$ coincides with 
the entanglement monotone called robustness of the entanglement for a bipartite pure state, where $D$ is the dimension of the bipartite Hilbert space $\Hi$.
In Section \ref{bipartite}, we show that the amount $D \beta_{\rightarrow } (\ket{\Psi})$ with one-way LOCC coincides with the Schmidt rank (the rank of the reduced density matrix) of the states. 
Also, as a corollary, we extend Cohen's characterization to a set of mixed states.
Finally, in section \ref{two-way}, constructing a concrete three-step two-way LOCC
discrimination protocol, we derive an upper bound for $\beta_{\leftrightarrow}(\rho) $. 
Calculating this upper bound analytically and also numerically, 
we show that $\beta_{\leftrightarrow}(\ket{\Psi})$ is
strictly smaller than $\beta_{\rightarrow } (\ket{\Psi} )$, and moreover, 
$\beta_{\rightarrow } (\rho )$ and $\beta_{\rm sep} (\rho)$ give
almost the same value for a lower dimensional bipartite pure state; 
this results can be seen in \textbf{ Figures 2,3,4,5,6}.   
As a result, we conclude that the two-way classical communication 
remarkably improves the local distinguishability in comparison with the one-way classical communication
for a low-dimensional pure state at least in the present problem settings.

%%%%%%%%%%%%%%%%%%%%%%%%%%%%%%%%%%%%%%%%%%%%%%%%%%%%%%%%%%%%%%

\section{Local discrimination  between an arbitrary state and the completely mixed state}\label{preliminary}
In this paper, we treat the bipartite system $\Hi := \Hi _A \otimes \Hi _B $ ($\dim \Hi = D$) composed of 
two finite-dimensional subsystems $\Hi _A$ and $\Hi _B$.
In the following sections, we often focus on the case when $\rho$ is pure.
In such a case, we assume that the dimension $d$ of $\Hi_A$ is equal to that
of $\Hi _B$. Note that the given pure state belongs to the composite system of the same-dimensional
subsystem. Then, the dimension ($D$) of the Hilbert space $\Hi$ is equal to $d^2$.
In the composite system $\Hi$, we call a positive operator $T$ with  $0 \le T \le I$ a one-way LOCC POVM element, where $I$ is an identity operator on $\Hi$, if the two-valued POVM $\{ T, I -T\}$ can be implemented by the one-way LOCC; 
we also define a two-way LOCC POVM element and a separable 
POVM element in the same manner by using the two-way LOCC and the separable operations
in stead of the one-way LOCC, respectively \cite{definition of LOCC}.
We write a set of one-way LOCC, two-way LOCC, separable POVM
elements, and all (global) POVM elements as $\mathcal{T} _{\rightarrow }$, $\mathcal{T}
_{\leftrightarrow}$, $\mathcal{T} _{\rm sep}$, and $\mathcal{T}_{\rm g}$. Obviously, they satisfy
the relation $\mathcal{T}_{\rightarrow} \subset
\mathcal{T}_{\leftrightarrow} \subset \mathcal{T}_{\rm sep} \subset \mathcal{T}_{\rm g}$. 
We can see that the condition $T \in \mathcal{T} _c$ is equivalent with the condition $I -T \in
\mathcal{T}_c$, where $c$ can be either $\rightarrow,
\leftrightarrow, {\rm sep}$, or ${\rm g}$.

In this paper, we discuss
the comparison of the performance of the local discrimination
in the case of the one-way LOCC, the two-way LOCC, and the separable operations.
In order to find this difference, although there are many problem
settings for the local discrimination, we especially focus on one
of the simplest problem settings as follows: We consider local
discrimination of an given arbitrary state $ \rho$ and 
another state $\tilde{\rho}$,
and investigate how well we can detect
$\tilde{\rho}$ under the additional condition that we do not make any
error to detect $\rho $
when the second state $\tilde{\rho}$ is the completely mixed state $\rho _{\rm mix} \stackrel{\rm def}{=} \frac{I_{AB}}{D } $ $(D=\dim \Hi)$;
namely, by only LOCC, how well we can distinguish
a given entangled state $\rho$ from the white noise state $\rho_{\rm mix}$
without making any error to judge the given state is $\rho _{\rm mix}$ when the real state is $\rho$.

Our problem can be written down rigorously as follows. 
We measure an unknown state
chosen from two candidates $\{ \rho , \tilde{\rho} \}$ by the two-values POVM $\{ T, I-T \}$, where
$T \in \mathcal{T} _{\rightarrow }, \mathcal{T} _{\leftrightarrow
},   \mathcal{T}_{\rm sep}, \mathrm{or} \  \mathcal{T}_{\rm g}$; that is, if we get the result
corresponding to $T$, then we decide that the unknown state is in $\rho$,
and if we get the result corresponding to $I-T$, then we decide
that the unknown state is in $\tilde{\rho}$. We consider two kinds of
error probability as follows: the type 1 error probability $\Tr
\rho (I-T)$, and the type 2 error probability $\Tr \tilde{\rho} T$; 
these are common terms in the field of ``{\it quantum hypothesis testing}'' \cite{Hayashi text}, 
where these two different error probabilities are treated in an asymmetric way.  
In this case, the type 1 error probability corresponds to the
error probability that the real state is $\rho$ and our decision
is $\tilde{\rho}$, and the type 2 error probability corresponds
to the error probability that the real state is $\tilde{\rho}$
and our decision is $\rho$. Thus, our problem is to
minimize the type 2 error probability $\Tr \tilde{\rho} T$
under the additional condition that the type 1 error probability $\Tr \rho
(I-T)$ must be $0$. Thus, we focus on the following minimum of the type
2 error probability:
\begin{equation}
\beta _c (\rho \| \tilde{\rho} ) := 
\min \{ \Tr (\tilde{\rho}
T) | T \in \mathcal{T}_c, \Tr \rho T = 1 \},
\end{equation}
where $c = \rightarrow$(one-way LOCC), $\leftrightarrow$(two-way LOCC), 
${\rm sep}$(separable operations), and ${\rm g}$(global operations).
When the both states $\rho$ and $\tilde{\rho}$ are pure states 
$|\Phi\rangle$ and $|\Psi\rangle$,
this quantity does not depend on whether $c = \rightarrow, \leftrightarrow, {\rm sep}, {\rm or} \ {\rm g}$, and is calculated as 
\begin{equation}
\beta _c (|\Phi\rangle \| |\Psi\rangle ) = |\langle \Phi|\Psi\rangle|^2
\end{equation}
for $c = \rightarrow$, $\leftrightarrow$, ${\rm sep}$, and ${\rm g}$.
This is because any discriminating protocol between two pure bipartite states
can be simulated by one-way LOCC 
when we focus only on the distribution of the outcome\cite{VSPM01, two-state-LD}.
In this paper, we focus on the minimum of the type 2 error probability 
in the case of $\tilde{\rho}=\rho_{\rm mix}$:
\begin{equation}
\beta _c (\rho ) :=\beta _c (\rho \| \rho_{\rm mix} )= \frac{t_c (\rho )}{D},
\label{normalization}
\end{equation}
where $t_c (\rho )$ is defined as
\begin{equation}\label{definition of t_c}
t_c(\rho ) = \min \{ \Tr T | T \in \mathcal{T}_c, \Tr T \rho = 1
\}.
\end{equation}
and $D$ is the dimension of the whole system $\Hi$.   
That is, $t_c (\rho)$ is in proportion to 
the minimum of the type 2 error probability 
$\beta _c (\rho)$ of one-way LOCC,
two-way LOCC, separable POVM and global POVM in the case where $c =
\rightarrow, \leftrightarrow$, ${\rm sep}$, and ${\rm g}$, respectively.
Trivially, 
\begin{equation}
t_{{\rm g}}(\rho ) = \rank \rho.
\end{equation}
Obviously, $t_c (\rho)$ satisfies the inequality $t_{{\rm g}}(\rho ) \le t_{{\rm sep}}(\rho ) \le
t_{\leftrightarrow}(\rho ) \le t_{\rightarrow}(\rho )$; as a matter of course, $\beta_c(\rho)$ also satisfies the similar inequality.
Note that by normalizing $\beta _c (\rho)$ as the above Eq.(\ref{normalization}),   
the resulting function $t_c(\rho )$ is no more a function depending both on $\rho$ and $\rho_{\rm mix }$,
but a function depending only on $\rho$. 

\begin{Remark}
In quantum information community, many papers treats the Bayesian framework,
in which the Bayesian prior distribution is assumed \cite{ACMMABV,NS,MP}.
However, in statistics community, non-Bayesian framework is more widely accepted,
in which no Bayesian prior distribution is assumed\cite{lehmann}.
This is because it is usually quite difficult to decide the Bayesian prior distribution based on the prior knowledge.
In order to resolve this difficulty,
they often treat the two kinds of error probabilities in an asymmetric way in hypothesis testing
without assuming prior distribution
because the importance of both error are not equal in a usual case, e.g., 
Neyman-Pearson lemma\cite{lehmann}, Stein's lemma\cite{Bucklew}, Hoeffding bound\cite{Hoe}.
These quantum cases are treated by several papers\cite{HP,Oga-Nag:test,Hoe-q,Nag-Hoe}.
In this paper, according to conventional statistics framework,
we focus on the error probabilities of the first and second,
and minimize the second kind of error probability under the constraint for the first one.
\end{Remark}

Here, we explain the reason why we choose the above special problem of discrimination of an arbitrary state $\rho$ from the completely mixed state $\rho_{\rm mix}$, 
and the reason why we add the above additional condition of perfect detection of $\rho$.
As we already said before, the first reason is that this additional condition makes the analysis of 
the problem extremely easier. 
Actually as we will see later in this paper,
we can derive the optimal POVM of this restricted local-discrimination problem with respect to each one-way LOCC and separable operations for a bipartite pure state.
As a result, we make the difference between one-way LOCC and two-way LOCC clear for our local-discrimination problem; this is our main purpose in this paper. 
Note that it is generally a hard problem to find an optimal protocol for a local-discrimination problem,  
and only in very limited situations, optimal local-discrimination protocols are known \cite{LD,VSPM01,two-state-LD}.
The second reason is that we can clearly see the relationship between local distinguishability and entanglement of a state in this problem setting.
In the previous paper \cite{local discrimination}, we showed the relationship between local distinguishability of a set of states 
and an average of the values of entanglement monotones for the states in terms of inequalities.
However, in this paper, we will show that the minimum error probability $\beta_c(\rho)$ of our problem is proportion to entanglement monotones in the case of
one-way LOCC ($c = \rightarrow $) and separable operations ($c = {\rm sep} $) at least for bipartite pure states except an unimportant constant factor.
The third reason is that the minimum error probability $\beta_c(\rho)$ can give a bound of local distinguishability 
for a more general local discrimination problem:
Suppose that a set of states $\{ \rho _i \}_{i=1}^{N_c}$
is perfectly locally distinguishable by one-way LOCC ($c = \rightarrow $),
two-way LOCC ($c=\leftrightarrow $), or separable ($c = \rm sep$)
POVM. From the result obtain in the the previous paper, $t_c (\rho
_i)$ (which corresponds to $d(\rho) $) gives an upper bound of $N
_c$ as \cite{local discrimination},
\begin{equation}\label{local dicrimination bound}
 N _c \le D/ \overline{t_c (\rho _i )} = 1/\overline{\beta_c(\rho _i)},
\end{equation}
where $\overline{t _c (\rho _i) } $ and $\overline{\beta_c(\rho _i)}$ are the average of $\{ t _c
(\rho _i) \}_{i=1}^{N_c}$ and $\{ \beta _c(\rho _i) \}_{i=1}^{N_c}$, respectively \cite{local discrimination}. 
Thus, $\beta_c(\rho )$ can be considered as an appropriate measure of local distinguishability in
a original operational sense, and also as a function whose average gives
an upper bound for the locally distinguishable sets of states.
Therefore, we investigate the difference of local
distinguishability of $\rho $ by one-way LOCC POVM, two-way LOCC
POVM, and separable POVM in terms of $\beta_c (\rho )$ in the
following sections.

%%%%%%%%%%%%%%%%%%%%%%%%%%%%%%%%%%%%%%%%%%%%%%%%%%%%%%%%%%%%%%%
\section{Local discrimination by separable POVM}\label{separable section chap.3}
In this section, we investigate the minimum type 2 error probability $\beta_{\rm sep}(\rho)= \frac{t_{{\rm sep}}(\rho)}{D}$ 
in terms of separable POVMs, which are given 
by $\{N_i\otimes M_i\}_i$ with the conditions $\sum_i N_i\otimes M_i=I$, $N_i\ge 0$, and $M_i \ge 0$.
The main purpose of this section is proving the following theorem: 
\begin{Theorem}\label{thm2}
The inequality
\begin{equation}\label{ineq-2}
t_{{\rm sep}}(\rho) \ge 
\max\{(\Tr \sqrt{\rho_A})^2,(\Tr \sqrt{\rho_B})^2\}
\end{equation}
holds for a bipartite state $\rho$ on $\Hi _A \otimes \Hi _B$, where $\rho _A$ and $\rho _B$ are the reduced density matrix of $\Hi _A$ and $\Hi _B$, respectively. 
Any pure state satisfies its equality.
In other words, the following inequality concerning the minimum error probability $\beta _{\rm sep}(\rho)$ holds:
\begin{equation}
\beta_{{\rm sep}}(\rho) \ge 
\frac{1}{D}\max\{(\Tr \sqrt{\rho_A})^2,(\Tr \sqrt{\rho_B})^2\}.
\end{equation}
\end{Theorem}
For a bipartite pure state,  the right-hand side of Eq.(\ref{ineq-2}) is proportional to
an entanglement monotone called the global robustness of entanglement $R_g(\ket{\Psi})$ except an unimportant constant term \cite{robustness paper}.

Applying Theorem \ref{thm2} for Eq.(\ref{local dicrimination bound}),
we can immediately derive the following corollary concerning the perfect discrimination of a given set of states in term of separable operations:
\begin{Corollary}
If a set of states $\{ \rho _i \}_{i=1}^N$ is perfectly distinguishable by separable operations, 
then, the set of states $\{ \rho _i \}_{i=1}^N$ satisfies the following inequality:
\begin{equation}\label{ineq-owari}
N \le D/\overline{\max \{ (\Tr \sqrt{\rho _{iA}})^2, (\Tr \sqrt{\rho_{iB}})^2 \}},
\end{equation}
where $\overline{\max \{ (\Tr \sqrt{\rho _{iA}})^2, (\Tr \sqrt{\rho_{iB}})^2 \}}$ is the average of 
$\max \{ (\Tr \sqrt{\rho _{iA}})^2, (\Tr \sqrt{\rho_{iB}})^2 \}$ for all $1 \le i \le N$.
\end{Corollary}
The above inequality is weaker than the inequality (\ref{local dicrimination bound}).
However, the inequality (\ref{ineq-owari}) is superior to the inequality (\ref{local dicrimination bound}) 
in terms of the efficiency of the computation;
that is, in general, we can not efficiently compute the bound in Eq.(\ref{local dicrimination bound}), 
since the function $t_{\rm sep}(\rho)$ includes the big variational problem.

\subsection{Pure states case}
First, for a technical reason, we 
concentrate the pure states case, and define a set of POVM elements
$\mathcal{T}_{\widetilde{\rm sep}}$ by,
\begin{eqnarray}
\mathcal{T} _{\widetilde{\rm sep}} \stackrel{\rm def}{=} \left \{ T \left|
\nonumber \quad T \le I_{AB}, T = \sum _i N _i \otimes M_i ,\quad
 \forall i, N_i \ge 0, M_i \ge 0 \right.\right \}.
\end{eqnarray}
$\mathcal{T} _{\widetilde{\rm sep}}$ is a set of POVM elements can be decomposed into a separable form; we say a positive
linear operator $M$ has a separable form, if $M/\Tr M$ is a
separable state. Since the definition of $\mathcal{T}
_{\widetilde{\rm sep}}$ is equivalent to the definition of $\mathcal{T}_{\rm sep}$
except the condition $I -T \in \mathcal{T}_{\rm sep}$, $\mathcal{T}_{\rm sep}$ is a subset of $\mathcal{T}_{\widetilde{\rm sep}} $. 
Note that even if $T \in \mathcal{T}_{\widetilde{\rm sep}}$, $I-T$ does not
necessary satisfy $I-T \in \mathcal{T}_{\widetilde{\rm sep}}$; that is,
$\mathcal{T}_{\rm sep}$ does not coincide with $\mathcal{T}_{\widetilde{\rm sep}} $. 
For example, suppose a set of states $\{ \ket{\Psi _i} \}_{i=1}^{m} \subset \Hi _A \otimes \Hi _B$ ($m < \dim \Hi $) is an unextendable product basis, 
and a POVM $T$ is defined as $T \stackrel{\rm def}{=} \sum _i \ket{\Psi _i}\bra{\Psi _i}$.
Then, $T$ belongs to $\mathcal{T}_{\widetilde{\rm sep}} $, but not to $\mathcal{T}_{\rm sep}$ 
since $I-T = I - \sum _{i=1}^m \ket{\Psi _i} \bra{\Psi _i} $ is proportion to a (bound) entangled state, 
and does not a separable form \cite{BDMSST99}.   
Similarly, we can define $t _{\widetilde{\rm sep}} $ as,
\begin{eqnarray}\label{definition of t widetilde}
t _{\widetilde{\rm sep}} (\rho ) \stackrel{\rm def}{=} \min \left \{ \Tr T |
 T \in \mathcal{T}_{\widetilde{\rm sep}}, \Tr \rho T = 1 \right \}.
\end{eqnarray}

 By definition, $t _{\widetilde{\rm sep}} (\rho )$ apparently
gives a lower bound of $t_{\rm sep} (\rho) = d^2\beta _{\rm sep}(\rho)$, that is, for all $\rho
\in S(\Hi)$,
\begin{equation}\label{inequality for t_sep }
t_{\widetilde{\rm sep}} (\rho) \le t_{\rm sep}(\rho) 
= d^2 \beta_{\rm sep}(\rho).
\end{equation}
Then, we can see that $t _{\widetilde{\rm sep}} (\rho) $ is actually
equal to $d(\rho)$ which is defined in Theorem 1 of the paper
\cite{local discrimination} as: 
\begin{equation}\label{definition of d}
d(\rho) \stackrel{\rm def}{=} \min \left \{ \frac{1}{\Tr \rho \omega} \Big| 0 \le \frac{\omega}{\Tr \rho \omega} \le I,
\omega \in {\rm SEP} \right \},
\end{equation}
where ${\rm SEP}$ is the set of all separable states.
We can easily check this fact just by defining $T \stackrel{\rm def}{=} \frac{\omega}{\Tr \rho \omega} $;
then, $T$ satisfies $0 \le T \le I$, $\Tr \rho T = 1$, and $T \in \mathcal{T}_{\widetilde{\rm sep}}$.
From Theorem 2 of the paper \cite{local discrimination}, 
for an arbitrary multipartite pure state $\ket{\Psi}$, 
$t _{\widetilde{\rm sep}}(\rho)$ satisfies the following 
inequality:
\begin{equation} \label{separable formula}
t_{\widetilde{\rm sep}} (\ket{\Psi } ) = d(\ket{\Psi}) \ge 1+
R_g(\ket{\Psi }),
\end{equation}
and $R_g(\ket{\Psi})$ is the global robustness of entanglement
\cite{robustness paper} defined as:
\begin{equation}\label{definition of robustness}
R_g(\rho) \stackrel{\rm def}{=} \min \left \{ t \ge 0 \Big| 
\exists {\rm a} \ {\rm state}\ \Delta, \ {\rm s.t.}\ \frac{1}{1+t}(\rho +t \Delta ) \in {\rm SEP}  \right \}.
\end{equation} 

For a bipartite pure state, we can know a more detail of $t_{\widetilde{\rm sep}} (\ket{\Psi } )$ as follows.
First, it was proven that $t_{\widetilde{\rm sep}} (\ket{\Psi } )$ coincides with 
the robustness of entanglement $R_g(\ket{\Psi}) $
for an arbitrary pure bipartite state $\ket{\Psi}$ \cite{Shash and Damian}. 
This fact can be seen by checking that the optimal states of $R_g(\rho)$, 
which was derived in \cite{robustness paper} satisfies the condition of
$d(\rho)$; the optimal state of $R_g(\rho)$ is also an optimal
state of $d(\rho )$.
Moreover, we know that the value of $R_g(\ket{\Psi})$ is given by the following formula for a bipartite state $\ket{\Psi}$ \cite{robustness paper}:
\begin{equation*}
R_g(\rho) = \Bigl(\sum _{i} \sqrt{\lambda _i}\Bigr)^2 -1, 
\end{equation*}
where $\{ \lambda _i \}_{i=1}^d$ is the Schmidt coefficients of $\ket{\Psi}$; $\ket{\Psi}$ can be decomposed as $\ket{\Psi} = \sum _i \sqrt{\lambda _i}
\ket{e_i}\otimes \ket{f_i}$ by choosing an appropriate orthonormal basis sets of the local Hilbert spaces $\{ \ket{e_i} \}_{i=1}^d \subset \Hi _A$ 
and $\{ \ket{f_i} \}_{i=1}^d \subset \Hi _B$.
Thus, we derive
\begin{Lemma}\label{lemma for bipartite t_widetilde}
For a bipartite pure state $\ket{\Psi} \in \Hi _A \otimes \Hi _B$,
\begin{equation}\label{bipartite t_widetilde}
t_{\widetilde{\rm sep}} (\ket{\Psi } ) = d(\ket{\Psi}) = 1+
R_g(\ket{\Psi }) = \Bigl(\sum _{i} \sqrt{\lambda _i}\Bigr)^2.
\end{equation}
\end{Lemma}
Although this lemma is a known result \cite{Shash and Damian}, as a preparation of the proof of the next theorem, 
we give a complete proof of Eq.(\ref{bipartite t_widetilde}),
in which we prove directly the equation $t_{\widetilde{\rm sep}} (\ket{\Psi } ) = \Bigl(\sum _{i} \sqrt{\lambda _i}\Bigr)^2$
from the definition of $t_{\widetilde{\rm sep}} (\ket{\Psi } )$.
\begin{Proof}
This proof is divided into two-steps. In the first step, we prove
that $\Bigl(\sum _{i} \sqrt{\lambda _i}\Bigr)^2$ is the lower bound of  $t_{\widetilde{\rm sep}}(\ket{\Psi})$. 
Then, in the second step, we construct POVM element $T$ which attains this lower
bound. For convenience, we define $\ket{M _{\Psi}} = \frac{1}{\sqrt{d}}\sum _{i=1}^d
\ket{e_i} \otimes \ket{f_i}$ where $\{ \ket{e_i} \}_{i=1}^d$ and $\{
\ket{f_i} \}_{i=1}^d$ are the Schmidt basis of $\ket{\Psi}$; thus, $\ket{M
_{\Psi}}$ is the maximum entangled state sharing the Schmidt
basis with $\ket{\Psi}$. Then, we derive $d | \braket{ M_{\Psi}
}{\Psi} |^2 = \Bigl(\sum _{i} \sqrt{\lambda _i}\Bigr)^2$.

 As the first step, we prove the following inequality;
\begin{eqnarray}
t_{\widetilde{\rm sep}} (\ket{\Psi})  &=& \min  \{ \Tr T | 0 \le T \le I,\ T \ {\rm is \ sep}, \ \bra{\Psi }T \ket{\Psi} =1 \} \nonumber \\
              &\ge & \min \{ d \bra{M_{\Psi}} T \ket{M_{\Psi}} | 0 \le T \le I, T \ {\rm is \ sep}, \bra{\Psi }T\ket{\Psi} =1 \} \label{1stineq}  \nonumber \\
              &\ge & d | \braket{ M_{\Psi} }{\Psi} |^2 \label{2ndineq}.
\end{eqnarray}
To prove the first inequality (\ref{1stineq}), since both $\Tr T$ and $d \bra{M_{\Psi}} T \ket{M_{\Psi}}$ are
linear for $T$, it is enough to prove only in the case that $T$ 
can be written down as 
$T =  \ket{a}\bra{a} \otimes \ket{b}\bra{b}$
by using un-normalized vectors $\ket{a}$ and $\ket{b}$.
Suppose $\ket{a} = \sum _{i=1}^d \alpha _i \ket{e_i}$ and $\ket{b} = \sum _{i=1}^d \beta _i \ket{f_i}$.
Then, using Schwarz's inequality, we can prove as follows
\begin{eqnarray*}
\Tr T = ( \sum _i | \alpha _i |^2)( \sum _i | \beta _i |^2) 
      \ge  |\sum _j \alpha _j \beta _j |^2 
      = d \bra{M_{\Psi}} T \ket{M_{\Psi}}.
\end{eqnarray*}
For the second inequality (\ref{2ndineq}),  since
the relations $\bra{\Psi}T\ket{\Psi} =1$ and $T \le I$ deduce that $\ket{\Psi}$
is an eigenvector of the largest eigenvalue $1$ of $T$, we derive $T \ge \ket{\Psi}
\bra{\Psi}$. Therefore, the inequality $d \bra{M_{\Psi}} T \ket{M_{\Psi}} \ge
d|\braket{M_{\Psi}}{\Psi} |^2$ is derived by taking the mean value with respect to $\ket{M
_{\Psi}}$.

As the second step, we construct an example of POVM element $T$
which achieves the lower bound we derived above. Define $T_0$ as
$T_0 = \ket{a} \bra{a} \otimes \ket{b}\bra{b}$ where $\ket{a} =
\sum _{i=1}^d (\lambda _i)^{1/4} \ket{e_i}$ and $\ket{b} = \sum _{i=1}^d
(\lambda _i)^{1/4}\ket{f_i}$; then, $T_0$ satisfies $\Tr T_0 =  
\Bigl(\sum _{i} \sqrt{\lambda _i}\Bigr)^2$. Moreover, since $P_0 T_0 P_0 =
\ket{\Psi} \bra{\Psi}$ where $P_0 = \sum _{i=1}^d \ket{e_i}
\bra{e_i} \otimes \ket{f_i} \bra{f_i}$, $T_0$ satisfies
$\bra{\Psi} T_0 \ket{\Psi} = 1$. 
Since $T_0$ apparently satisfies $0 \le T_0$, the inequality $T \le I$ is the only remaining
condition which the optimal POVM element $T$ attaining the equality $\Tr T = t_{\widetilde{\rm sep}} (\ket{\Psi})$ must satisfy. 
Since $T_0$ does not generally satisfies the inequality $T_0 \le I$,
we construct a new POVM element $T$ which satisfies $0 \le T \le I$ from $T_0$. 
In order to
construct the POVM element $T$ from $T_0$, we use twirling technique here. We define a
family of local unitary operators $U_{\overrightarrow{\theta}}$ parameterized by
$\overrightarrow{\theta} = \{\theta _i \}_{i=1}^d$ as follows,
\begin{equation}\label{definition of U_theta}
U_{\overrightarrow{\theta}} = (\sum_{j=1}^d e^{i \theta _j} \ket{e_j}\bra{e_j})
\otimes (\sum _{k=1}^d e^{-i \theta _k} \ket{f_k}\bra{f_k} ).
\end{equation}
Note that $\left (\Hi ^{\otimes 2},  U_{\overrightarrow{\theta}}  \right )$ 
is a unitary representation of the compact topological group $\overbrace{U(1) \times \cdots \times U(1)}^n $;
by means of a unitary representation of a compact topological group, 
we implement the "{\it twirling}" operation (the averaging  over the compact topological group) for a state (or POVM) \cite{HMMOS07}.    
Then, we define $T$ as the operator which is constructed by
twirling $U_{\overrightarrow{\theta}} T_0 U_{\overrightarrow{\theta}}^{\dagger}$ over parameters $\overrightarrow{\theta} = \{
\theta _i \}_{i=1}^d$. 
Since by an action of twirling operation, a given state is projected to the subspace of all invariant elements of the group action \cite{HMMOS07},
we can calculate $T$ as follows:
\begin{eqnarray*}
T & \stackrel{\rm{ def} }{=} & \int  _0^{2 \pi} \cdots \int _0^{2
\pi} U_{\overrightarrow{\theta}} T_0 U_{\overrightarrow{\theta}}^{\dagger} d \theta _1 \cdots d \theta _d \\
&=& (\sum _{j=1}^d \ket{e_j}\bra{e_j} \otimes \ket{f_j}\bra{f_j} )T_0 ( \sum _{j=1}^d \ket{e_j}\ket{e_j} \otimes \ket{f_j}\bra{f_j}) \\
&\ & \quad +\sum _{j \neq k} \left ( \ket{e_j}\bra{e_j}\otimes \ket{f_k}\bra{f_k} \right ) T_0 \left ( \ket{e_j}\bra{e_j}\otimes \ket{f_k}\bra{f_k} \right ) \\ 
& = & (\sum _{i=1}^d \sqrt{\lambda _i} \ket{e_i} \otimes \ket{f_i})
 (\sum _{i=1}^d \sqrt{\lambda _i} \bra{e_i} \otimes \bra{f_i} ) \nonumber \\
 & \ & \quad + \sum _{i \neq j} \sqrt{\lambda _i \lambda _j}  \ket{e_i}\bra{e_i}
 \otimes \ket{f_j}\bra{f_j}.
\end{eqnarray*}
Since $\sqrt{\lambda _i \lambda _j} \le 1$, $T \le I$.  Moreover, $T$ satisfies $0
\le T \le I$, $\bra{\Psi} T \ket{\Psi} =1$, and is in the separable
form; we only applied the local unitary $U_{\overrightarrow{\theta}}$ to un-normalized product state $T_0$, and, 
then, took an average over parameters $\overrightarrow{\theta}$. Thus, we derive the inequality 
$t_{\widetilde{\rm sep}}(\ket{\Psi}) \le \Tr T = \Bigl(\sum _{i} \sqrt{\lambda _i}\Bigr)^2$.
Since we have already proven the converse inequality, 
we conclude $t_{\widetilde{\rm sep}}(\ket{\Psi}) = \Bigl(\sum _{i} \sqrt{\lambda _i}\Bigr)^2$ \\ \hfill $\square$
\end{Proof}
Finally, by means of Lemma \ref{lemma for bipartite t_widetilde}, we can derive the following theorem, i.e., Theorem \ref{thm2} in the pure states case:
\begin{Theorem}\label{theorem separable main}
For a bipartite pure state $\ket{\Psi}$,
\begin{eqnarray}\label{main theorem of sep POVM}
\fl  \beta_{\rm sep} (\ket{\Psi } )=\frac{1}{d^2}t_{\rm sep}(\ket{\Psi})  &=& \frac{1}{d^2}\left (1+
R_g(\ket{\Psi }) \right) \nonumber \\
&=&   \frac{1}{d^2}\Bigl(\sum _{i} \sqrt{\lambda _i}\Bigr)^2 = \frac{1}{d^2}(\Tr \sqrt{\rho _A})^2 = \frac{1}{d^2}(\Tr \sqrt{\rho _B})^2,
\end{eqnarray}
where $\{ \lambda _i \}_{i=1}^d$ is the Schmidt coefficients of $\ket{\Psi}$.
\end{Theorem}
\begin{Proof}
Since by the definition $t_{\widetilde{\rm sep}} (\ket{\Psi } ) \le t_{\rm sep} (\ket{\Psi } ) $,
all what we need to prove is that 
the optimal POVM $T$ for $t_{\widetilde{\rm sep}} (\ket{\Psi } )$ is also the optimal POVM for $t_{\rm sep} (\ket{\Psi } )$;
that is, $I -T$ also has a separable form.

As we have already shown in the proof of Lemma \ref{lemma for bipartite t_widetilde}, 
the optimal POVM $T$ for $t_{\widetilde{\rm sep}} (\ket{\Psi } )$ can be written down as 
\begin{equation}\label{optimal POVM element separable}
\fl T = (\sum _{i=1}^d \sqrt{\lambda _i} \ket{e_i} \otimes \ket{f_i})
 (\sum _{i=1}^d \sqrt{\lambda _i} \bra{e_i} \otimes \bra{f_i} ) 
~ + \sum _{i \neq j} \sqrt{\lambda _i \lambda _j}  \ket{e_i}\bra{e_i}
 \otimes \ket{f_j}\bra{f_j},
\end{equation}
where $\{ \ket{e_i}\otimes \ket{f_j} \}_{ij}$ and $\{ \lambda _i \}_{i=1}^d $ are the Schmidt basis and the Schmidt coefficients
corresponding to $\ket{\Psi} = \sum _{i=1}^d \sqrt{\lambda _i} \ket{e_if_j}$, respectively.
Suppose 
\begin{equation*}
\fl \overline{T_0} \stackrel{\rm def}{=} \frac{1}{2}\sum _{i\neq j}\ket{\overline{a_{ij}}}\bra{\overline{a_{ij}}} 
\otimes \ket{\overline{b_{ij}}}\bra{\overline{b_{ij}}}
+ \sum _{i\neq j}\{ \sum _{k \neq i,j} \lambda _k + (\sqrt{\lambda _i} - \sqrt{\lambda _j})^2\}\ket{e_if_j} \bra{e_if_j},
\end{equation*}
where $\ket{\overline{a_{ij}}}$ and $\ket{\overline{b_{ij}}}$ are defined as
$\ket{\overline{a_{ij}}} \stackrel{\rm def}{=} (\lambda _j)^{\frac{1}{4}}\ket{e_i}-(\lambda _i)^{\frac{1}{4}}\ket{e_j}$
and
$\ket{\overline{b_{ij}}} \stackrel{\rm def}{=} (\lambda _j)^{\frac{1}{4}}\ket{f_i}+(\lambda _i)^{\frac{1}{4}}\ket{f_j}$ for $i \neq j$,
respectively.
Then, as is proven in \ref{appendix}, the relation
\begin{equation}
\int  _0^{2 \pi} \cdots \int _0^{2
\pi} U_{\overrightarrow{\theta}} \overline{T_0} U_{\overrightarrow{\theta}}^{\dagger} d \theta _1 \cdots d \theta _d
=I-T
\label{average}
\end{equation}
holds, where a local unitary operator $U_{\overrightarrow{\theta}}$ is defined as Eq.(\ref{definition of U_theta}).
By the definition, $\int  _0^{2 \pi} \cdots \int _0^{2
\pi} U_{\overrightarrow{\theta}} \overline{T_0} U_{\overrightarrow{\theta}}^{\dagger} d \theta _1 \cdots d \theta _d$ is apparently
a separable POVM element. Therefore, we can conclude the equality $t_{\widetilde{\rm sep}} (\ket{\Psi } ) = t_{\rm sep} (\ket{\Psi } ) $ for a bipartite pure state.
By means of Lemma \ref{lemma for bipartite t_widetilde}, 
we derive Eq.(\ref{main theorem of sep POVM}). \hfill $\square$
\end{Proof}

Thus, $t_{\rm sep} (\ket{\Psi } )$ is equivalent to $1+ R_g(\ket{\Psi})$, and
the minimum type 2 error probability $\beta_{\rm sep}(\ket{\Psi}) $ only depends on the global 
robustness of entanglement $R_g(\ket{\Psi})$ for a bipartite pure state $\ket{\Psi}$.
In this case, the optimal POVM $\{ T, I-T \}$ can be derived by using Eq.(\ref{optimal POVM element separable}) as the definition of the POVM element $T$. 

We should note that Theorem \ref{theorem separable main} 
not only gives a way to calculate the minimum type 2 error probability under separable operations 
$\beta_{\rm sep}(\ket{\Psi})$,
but this theorem gives a complete 
relationship between the local distinguishability of 
a bipartite state under separable operations 
and the entanglement of the state. In the previous paper \cite{local discrimination}, 
it was shown the global robustness of entanglement $R_g(\ket{\Psi})$ gives an upper-bound for the maximum number of 
distinguishable states under separable operations. 
However, the present result shows that $R_g(\ket{\Psi})$ is nothing but the local distinguishability (against the completely mixed state) itself
at least for a bipartite pure state. 
In other words, it is shown that robustness of entanglement has rigorously operational meaning for bipartite pure states
in terms of the local discrimination from the completely mixed state
$\rho_{\rm mix}$.

\subsection{Mixed states case}
Now, we prove Theorem \ref{thm2} for a general mixed bipartite state.
\begin{Proof}[Theorem \ref{thm2}]
First we prove the inequality 
$t_{\rm sep}(\rho)\ge (\Tr \sqrt{\rho_A})^2$.
Adding the system $B'$, we choose a purification $|\Phi\rangle$ of $\rho$.
In the following, we will prove the inequality 
$t_{\widetilde{\rm sep}}(\rho) \ge t_{\widetilde{\rm sep}}(\ket{\Phi})$.
If this inequality holds, 
applying Eq.(\ref{inequality for t_sep }) and Eq.(\ref{bipartite t_widetilde}), we obtain
$t_{\rm sep}(\rho)\ge t_{\widetilde{\rm sep}}(\rho) \ge t_{\widetilde{\rm sep}}(\ket{\Phi}) = (\Tr \sqrt{\rho_A})^2$.

Define a separable positive operator 
$T=\sum_{i}p_i |e_i\rangle\langle e_i|
\otimes |f_i\rangle\langle f_i|$ 
$(e_i \in \Hi_A, f_i \in \Hi_B, \|f_i\|=1, \|e_i\|=1)$
such that $0 \le T \le I_{AB}$ and $\Tr \rho T=1$.
Thus, 
$\langle \Phi| \sum_{i}p_i 
\bigl(|e_i\rangle\langle e_i|\otimes |f_i\rangle\langle f_i|\otimes I_{B'} \bigr)
|\Phi\rangle=1$.
Now, we focus on the bipartite system $A$ and $BB'$.
Then, we choose the state $|\tilde{f}_i\rangle
$ $(\|\tilde{f}_i\|=1)$ on $\Hi_{BB'}$
such that
$\Tr_{A} (|\Phi\rangle\langle \Phi |) 
(|e_i\rangle\langle e_i|\otimes I_{BB'})=
c_i |\tilde{f}_i\rangle\langle \tilde{f}_i|$,
where $c_i$ is the normalizing constant.
Define the state $|f_i'\rangle
$ $(\|f_i'\|=1)$ on $\Hi_{BB'}$ by
\begin{equation}
|f_i'\rangle \langle f_i'|
:= 
\frac{1}{\langle \tilde{f}_i|P_i|\tilde{f}_i\rangle }
P_i |\tilde{f}_i\rangle\langle \tilde{f}_i|P_i 
\le P_i,
\end{equation}
where the projection $P_i$ is defined by 
$P_i := |f_i\rangle\langle f_i|\otimes I_{B'}$.
Since $\langle \tilde{f}_i| P_i |\tilde{f}_i\rangle
= \langle \tilde{f}_i|f_i'\rangle \langle f_i'|\tilde{f}_i\rangle
$,
\begin{eqnarray*}
\fl \Tr \rho  (|e_i\rangle\langle e_i|\otimes |f_i\rangle\langle f_i|)
=\langle \Phi|( |e_i\rangle\langle e_i|
\otimes |f_i\rangle\langle f_i|\otimes I_{B'}|)|\Phi\rangle
=\langle \Phi| (|e_i\rangle\langle e_i|
\otimes |f_i'\rangle\langle f_i'|)|\Phi\rangle.
\end{eqnarray*}
Thus, the relations
\begin{eqnarray*}
T':= \sum_{i} p_i |e_i\rangle\langle e_i|
\otimes |f_i'\rangle\langle f_i'|
\le 
\sum_{i} p_i |e_i\rangle\langle e_i|
\otimes |f_i\rangle\langle f_i| \otimes I_{B'}
\le I_{ABB'}\\
\langle \Phi |T'|\Phi\rangle=\Tr \rho T=1
\end{eqnarray*}
hold. 
Moreover, $T'$ satisfies the equality $\Tr T' = \sum_{i} p_i = \Tr T$.
Thus, the inequality $t_{\widetilde{\rm sep}}(\rho) \ge t_{\widetilde{\rm sep}}(\ket{\Phi})$ holds.
Therefore, the relations 
$t_{\rm sep}(\rho)\ge t_{\widetilde{\rm sep}}(\rho) \ge t_{\widetilde{\rm sep}}(\ket{\Phi}) = (\Tr \sqrt{\rho_A})^2$ hold.

Similarly, we can show 
the inequality 
$t_{\rm sep}(\rho)\ge (\Tr \sqrt{\rho_B})^2$.
Thus, we obtain (\ref{ineq-2}) in the mixed states case.
\hfill $\square$
\end{Proof}

%%%%%%%%%%%%%%%%%%%%%%%%%%%%%%%%%%%%%%%%%

%%%%%%%%%%%%%%%%%%%%%%%%%%%%%%%%%%%%%%%%%%%%%%%%%%%%%%%%%%%%%%%%%%%

\section{Local discrimination by one-way LOCC} \label{bipartite}
In this section, we 
prove the following theorem concerning 
the local discrimination problem in terms of one-way LOCC in the direction $A \to B$:
\begin{Theorem}\label{theorem one-way main}
The inequality 
\begin{equation}
\label{ineq-1}
t_{\rightarrow}(\rho)
\ge \rank \rho_A
\end{equation}
holds for a bipartite state $\rho $ on $\Hi _A \otimes \Hi _B$.
Any maximally correlated state $\rho$ satisfies the equality.
In other words, the following inequality concerning the minimum error probability $\beta _{\rightarrow}(\rho)$ holds:
\begin{equation}
\beta_{\rightarrow}(\rho)
\ge \frac{1}{D}\rank \rho_A
\end{equation}
\end{Theorem}
In the above theorem, a maximally correlated state is defined as a state which can be decompose in the following form:
\begin{equation}\label{def of maximally correlated} 
\rho = \sum _{1 \le i,j \le d} \alpha _{ij}\ket{u_i, v_i }\bra{u_j, v_j},
\end{equation}
where $\{ \ket{u_i} \}_{i=1}^d$ and $\{ \ket{v_j} \}_{j=1}^d$ are orthonormal bases of $\Hi _A$ and $\Hi _B$, respectively \cite{maximally correlated state}; 
apparently, a pure state is a maximally correlated state.
Thus, $t_{\rightarrow}(\ket{\Psi})=D \beta_{\rightarrow }(\ket{\Psi} )$ 
is equal to the Schmidt rank of a state for a bipartite pure state $\ket{\Psi}$.
In the case when $\rho$ is a maximally correlated state satisfying Eq.(\ref{def of maximally correlated}), the optimal way to discriminate between $\rho$ and the completely mixed state is
the following: Suppose there are two parties called Alice and Bob. Both Alice and Bob measure their local states $\Hi _A$ and $\Hi _B$ in
the bases $\{ \ket{u_i} \}_{i=1}^d$ and $\{ \ket{v _j} \}_{j=1}^d$, respectively, (of course, they only need to detect the
support of the local states). Then, Alice informs her
measurement result to Bob. Suppose Alice's result is $\ket{u_k}$ and Bob's result is $\ket{v_l}$.
If $k$ is equal to $l$, then, they judge that the given state is $\rho$. Otherwise, they judge that the given state is the completely mixed state.

By comparing Theorem \ref{thm2} and Theorem \ref{theorem one-way main},
we can easily see that if a bipartite pure state
$\ket{\Psi}$ is not a maximally entangled state nor a product
state, then, the strict inequality $ \beta_{\rm sep}
(\ket{\Psi }) < \beta_{\rightarrow } (\ket{\Psi })$ holds. Thus, 
we can conclude that  there is a gap between the
one-way local distinguishability and the separable local
distinguishability for a bipartite pure state at least in the present problem settings
from these results.

Applying Theorem \ref{theorem one-way main} for Eq.(\ref{local dicrimination bound}),
we can extend 
Cohen's characterization \cite{cohen} concerning the perfect discrimination of a given set of pure states in term of one-way LOCC
to a set of mixed states:
\begin{Corollary}
If a set of bipartite states $\{ \rho _i \}_{i=1}^N$ on $\Hi _A \otimes \Hi _B$ is perfectly distinguishable by one-way LOCC,
then,  
\begin{equation}\label{ineq-one-way}
\sum _{i=1}^N {\rm rank} \rho _{iA} \le D.
\end{equation}
\end{Corollary}
This bound of the size of locally distinguishable sets for one-way LOCC is much stronger than the known bound for separable operations 
\cite{local discrimination}.

As a preparation for our proof of Theorem \ref{theorem one-way main}, 
we see the fact that there are several equivalent representations of
the definition of one-way LOCC POVM elements. We start from the
following representation which we can see immediately from
the definition; that is, in a bipartite system $\Hi = \Hi _A
\otimes \Hi_B$ , if $T \in \mathcal{T} _{\rightarrow } $, there
exist sets of positive operators $\{ M_i \}_i $ and $\{ N_j^i \}_j
$ such that
\begin{equation} \label{first definition}
T =\sum _{ij} M_i \otimes N _j^i,
\end{equation}
$\sum _i M_i \le I_{B}$, and $\sum _j N_j^i \le I_A$, where $\{ M
_i \}_i$ is the POVM of the local measurement on $\Hi_A$ and $\{
N_j^i \}_j$ is the POVM of the local measurement on $\Hi_B$
depending on the first measurement result $i$. 
Further, redefining $N_j^i $ as ${N}_i=
\sum _j N _j^i$, we derive the following 
equivalence relation:
\begin{eqnarray}
& \quad & T \in \mathcal{T}_{\rightarrow} \nonumber \\
& \Longleftrightarrow & \exists \{ M_i \} _i \ \mathrm{and}
\nonumber \ \{ N _i \} _i \\ &\quad & s.t. \forall i, 0 \le M _i ,
0 \le N_i \le I_B, \sum _i M _i \le I_A,
 \mathrm{and } \ T = \sum _{i} M_i \otimes N _i .\label{first formulation}
\end{eqnarray}
Using this characterization, we obtain 
the following lemma.
\begin{Lemma} \label{probability}
A one-way LOCC POVM element $T = \sum _{ij} M_i\otimes N_j^i  \in
\mathcal{T}_{\rightarrow}$ satisfies $\Tr \rho T =1$ if and only
if  $\Tr ( \rho _A \sum M_i ) = 1$ and $\Tr (\rho _{B, M_i} \sum
_j N_j^i ) = 1$ for all $i$, where $\rho _A \stackrel{\rm def}{=}
\Tr _B \rho$ and $\rho _{B, M_i} \stackrel{\rm def}{=} \Tr _A \rho
M_i \otimes I _B / \Tr \rho M_i \otimes I _B$.
\end{Lemma}
\begin{Proof}
We can calculate $\Tr \rho T$ as follows:
\begin{eqnarray*}
\Tr \rho T = \sum _{ij} \Tr \rho M_i \otimes N_j^i
&=& \sum _{ij} \Tr \{ (\Tr _A \rho (M_i \otimes I_B) ) N_j^i \} \\
&=& \sum _i \Tr \rho _A M_i \cdot \Tr \rho _{B , M_i } (\sum _j
N_i^j) =1.
\end{eqnarray*}
Since $\sum _i \Tr \rho _A M_i \le 1$ and $\Tr \rho _{B, M_i}
(\sum _j N_j^i) \le 1$ for all $i$, we derive $\sum _i \Tr \rho _A
M_i = 1$ and $\Tr \rho _{B, M_i} (\sum _j N_j^i) = 1$. The
opposite direction is trivial.
 \hfill $\square$
\end{Proof}

Now, we prove Theorem \ref{theorem one-way main}
using the above lemma.
\begin{Proof}[Theorem \ref{theorem one-way main}]
In order to detect a state perfectly, we need to
detect the reduced density operator of the local system $A$, $\rho
_A$ as well as that of the other local system $B$, $\rho _{B,M_i}$, 
perfectly in each step.
Thus, we can assume that $N_i$ is a projection on $B$ 
without loss of generality.
Hence, $\Tr T = \sum_i \Tr M_i \cdot \Tr N_i \ge \sum_i \Tr M_i$.
Since we have to detect the reduced density operator 
of the local system $A$, $\rho_A$,
we obtain $\Tr \sum_i M_i\rho_A= 1$, i.e., (\ref{ineq-1}).
When the state $\rho$ is a maximally mixed state
$\sum_{1\le i,j\le d} a_{i,j}|u_i,v_i\rangle \langle u_j,v_j|$,
the reduced density $\rho_A$ is 
$\sum_{i=1}^d a_{i,i}|u_i\rangle \langle u_i|$.
Thus, $\rank \rho_A= d$.
In this case, we can perfectly detect this state by 
the one-way LOCC test
$\sum_{i=1}^d |u_i\rangle \langle u_i | \otimes |v_i\rangle \langle v_i|$.
\hfill $\square$
\end{Proof}

We should note the following fact: Although a maximally correlated state satisfies the equality of Eq.(\ref{ineq-1}), 
the converse is not necessarily true.
Even if $v_i$ is not orthogonal, 
we can perfectly detect this state by 
the one-way LOCC test
$\sum_{i=1}^d |u_i\rangle \langle u_i | \otimes |v_i\rangle \langle v_i|$.
When the rank of the state 
$\sum_{1\le i,j\le d} a_{i,j}\braket{v_j}{v_i} |u_i\rangle \langle u_j|$
is $d$,
the rank of $\rho_A$ is $d$.
That is, this gives a counter example of the converse.

%%%%%%%%%%%%%%%%%%%%%%%%%%%%%%%%%%%%%%%%%%%%%%%%%%%%%%%%%%%%%%%%%%

%%%%%%%%%%%%%%%%%%%%%%%%%%%%%%%%%%%%%%%%%%%%%%%%%%%%%%%%%%%
\section{Local discrimination by two-way LOCC} \label{two-way}
So far, we have calculated the minimum error probability of the local discrimination problem
for one-way LOCC $\beta_{\rightarrow}(\ket{\Psi})$,
and separable operations $\beta_{\rm sep}(\ket{\Psi})$.
In this section, we focus on discrimination protocols by two-way LOCC.
Since the two-way LOCC is mathematically complicated, 
it is difficult to derive the minimum two-way LOCC discrimination protocol, and as a result,
it is difficult to derive the exact value of $\beta_{\leftrightarrow}(\ket{\Psi })$.
However, in order to show the difference of the efficiency of one-way and two-way local discrimination protocols, 
(which is actually our main purpose of this paper,)
it is enough to find the upper-bound of the two-way error probability $\beta_{\leftrightarrow}(\ket{\Psi })$.
Thus, we concentrate ourselves on driving an upper-bound of $\beta_{\leftrightarrow}$
by constructing a concrete two-way LOCC discrimination protocol.
For simplicity, we only treat three-step LOCC discrimination protocols on a bipartite system, 
which are in the simplest   class of genuine two-way LOCC protocols. 
As a result, we show that even three-step LOCC protocols can discriminates a given state
from the completely mixed state much better than by one-way (that is, two-step) LOCC protocols.

We can generally describe a three-step LOCC protocol to discriminate a pure state $\ket{\Psi }$ 
from $\rho _{\rm mix} = \frac{I}{d^2}$ on a bipartite system $\Hi _A \otimes \Hi _B$ 
without making any error to detect $\ket{\Psi }$ as follow: 
Suppose there are two parties called Alice and Bob. First, Alice performs a POVM $\{ M _i \}_i$ on her system $\Hi _A$, and sends
the measurement result $i$ to Bob. Second, depending on $i$, Bob performs a POVM $\{ N _j^i \}_j$ on his system $\Hi _B$, and
sends the measurement result $j$ to Alice. If the given state is $\rho$, by easy calculation, 
we can check that the Alice's state after this step is 
\begin{equation}\label{definition of sigma _A}
\sigma _A^{ij} \stackrel{\rm def}{=} 
\frac{\sqrt{M_i}\sqrt{\rho _A}N_j^{iT}\sqrt{\rho _A}\sqrt{M_i}}{  \Tr \left (  M_i \sqrt{\rho _A}N_j^{iT}\sqrt{\rho _A} \right) },
\end{equation}
where $\rho _A \stackrel{\rm def}{=} \Tr _B \ket{\Psi }\bra{\Psi }$, and the transposition is taken in the Schmidt basis of $\ket{\Psi }$.
Thus, in order not to make an error to detect the above state, finally, 
Alice should make a measurement in $\{ \{ \sigma _A^{ij} > 0 \},  I_A - \{ \sigma _A^{ij} > 0\} \}$,
where $\{ \sigma _A^{ij} > 0\}$ is a projection operator onto the support of $\sigma _A^{ij}$ 
(the subspace spanned by eigenvectors corresponding to non-zero eigenvalue of $\sigma _A^{ij}$ ), 
$I_A$ is an identity operator in $\Hi _A$. 
Then, if she detects $\{ \sigma _A^{ij} > 0 \} $, she judges that the first state was $\ket{\Psi }$, 
and if she detects $I_A - \{ \sigma _A^{ij} > 0\}$, she judges that the first state was $\rho _{\rm mix}$. 
Suppose $\{ T, I_{AB} - T\}$ is the POVM corresponds to the above local discrimination protocol,
where $T$ corresponds to $\ket{\Psi }$, and $I -T $ corresponds to $\rho _{\rm mix}$
Then, we can check that the whole POVM $\{ T, I-T \}$ can be written down as follows,
\begin{equation}\label{definition of two-way T}
T = \sum _i \sum _j  \left ( \sqrt{ M_i} \{ \sigma_A^{ij} >0 \} \sqrt{ M_i} \right ) \otimes N_j^i,
\end{equation}
where $\sigma _A^{ij}$ is defined by Eq.(\ref{definition of sigma _A}), 
and all $ M_i$ and $N_j^i$ are positive operators satisfying $\sum _i M_i = I_A$ and $\sum _j N_j^i =I_B$. 
We can also check that $T$ defined by Eq.(\ref{definition of two-way T}) satisfies $\bra{\Psi}T\ket{\Psi} = 1$ as follows:
\begin{eqnarray*}
\fl \bra{\Psi}T\ket{\Psi} &=& \sum _{ij} \bra{\Psi}\left ( \sqrt{M_i } \{ \sigma_A^{ij} >0 \} \sqrt{M_i } \right ) \otimes N_j^i\ket{\Psi} \\
&=& d \sum _{ij} \left ( \bra{\Phi ^+}\sqrt{\rho _A} \otimes I_B \right ) \left (\sqrt{M_i } \{ \sigma_A^{ij} >0 \} \sqrt{M_i } \right ) \otimes N_j^i
\left ( \sqrt{\rho _A} \otimes I_B \ket{\Phi^+} \right ) \\
&=& d \sum _{ij} \bra{\Phi ^+} \left ( \sqrt{N_j^i}^T \sqrt{\rho _A}\sqrt{M_i } \{ \sigma_A^{ij} >0 \} \sqrt{M_i } \sqrt{\rho _A} \sqrt{N_j^i}^T \right ) 
\otimes I_B \ket{\Phi ^+} \\
&=& \sum _{ij} \Tr \sqrt{N_j^i}^T \sqrt{\rho _A}\sqrt{M_i } \{ \sigma_A^{ij} >0 \} \sqrt{M_i } \sqrt{\rho _A} \sqrt{N_j^i}^T \\
&=& \sum _{ij} \Tr \sqrt{M_i } \sqrt{\rho _A}N_j^{iT}\sqrt{\rho _A}\sqrt{M_i } \{ \sigma_A^{ij} >0 \} \\
&=& \sum _{ij} \left ( \Tr M_i  \sqrt{\rho _A}N_j^{iT}\sqrt{\rho _A} \right ) \cdot \left ( \Tr \sigma _A^{ij} \{ \sigma_A^{ij} >0 \} \right )\\
&=& \sum _{ij} \Tr \left ( N_j^{i}\left ( \sqrt{\rho _A} M_i  \sqrt{\rho _A} \right )^T \right )\\
&=& \sum _{ij} \Tr \rho _A M _i\\
&=& 1,
\end{eqnarray*}
where $\ket{\Phi ^+}$ is the maximally entangled state sharing the Schmidt basis with $\ket{\Psi}$,
and the transposition $T$ is always taken in the Schmidt basis of $\ket{\Psi }$.
In the second line of the above equalities, 
we used the equality $\ket{\Psi } = \sqrt{\rho _A} \otimes I_B \ket{\Phi ^+}$.
In the third line, we used the equalities $I _A \otimes X \ket{\Phi ^+} = X^T \otimes I_B \ket{\Phi ^+}$, which is valid for an arbitrary operator $X$.
In the sixth line, we used Eq.(\ref{definition of sigma _A}).

The above three-step LOCC protocol is enough general. 
However, it is too complicated to optimize $\Tr T$ over all choices of POVM $\{ M _i \}_i$ and $\{ N_j^i \}_j$. 
In this section, our aim is only to find a good (not necessary optimal) two-way LOCC protocol by 
which we can discriminate a state from the completely mixed state better than by any one-way LOCC protocols.
Thus, to make a problem simpler, we make the following assumptions on Alice's POVM $\{ M _i \}_i$ and Bob's POVM $\{ N_j^i \}_j$:
First, without losing any generality, 
we can write $\ket{\Psi }$ as $\ket{\Psi } = \sum _{k=1}^d \sqrt{\lambda _k} \ket{k}\otimes \ket{k}$ with $\lambda _i > 0$ and $\lambda _i \ge \lambda _{i+1}$,
where $\{ \ket{k} \}_{k=1}^d$ is an arbitrary fixing computational basis of $\Hi _A$ and $\Hi _B$, and $d = \dim \Hi _A = \dim \Hi _B$;
since our definition of $t_{\leftrightarrow }(\ket{\Psi})$ (Eq.(\ref{definition of t_c})) 
does not depend on the dimension of the whole system, 
we can always choose the whole system in order that the Schmidt rank of $\ket{\Psi}$ is $d$.
Second, we assume that the number of POVM element $M_i$ is $d $, and 
$M_i$ is diagonalizable in the computational basis $\{ \ket{k} \}_{k=1}^d$ as,
\begin{equation}
M_i = \sum _{k=1}^{i} \delta _{ki}\ket{k}\bra{k}, 
\end{equation}  
where ${\rm rank} M_i = i$, and the coefficients $\delta _{ki} \ge 0$ satisfy $\sum _{i=k}^d \delta _{ki} = 1$ for all $k$.
Moreover, we assume that $\{ N_j^i \}_j$ is a von Neumann measurement corresponding to a mutually unbiased basis 
\cite{VSPM01, WF89} $\{ \ket{\xi _j^i} \}_{j=1}^{{\rm rank} M_i}$
of an orthonormal set of eigen vectors of $\omega _B \stackrel{\rm def}{=} \frac{\sqrt{\rho _B}M_i^T \sqrt{\rho _B}}{\Tr \sqrt{\rho _B}M_i^T \sqrt{\rho _B}}$
corresponding to non-zero eigenvalues; that is, $\{ \xi _j^i \}_{j=1}^{{\rm rank} M_i}$ only spans 
 ${\rm Ran } \frac{\sqrt{\rho _B}M_i^T \sqrt{\rho _B}}{\Tr \sqrt{\rho _B}M_i^T \sqrt{\rho _B}}  $. 
In other words, an orthonormal set of states $\{ \ket{\xi _j^i} \}_{j=1}^{{\rm rank} M_i}$ 
satisfies 
\begin{equation}\label{mutually unbiased}
\bra{\xi _j^i}\frac{\sqrt{\rho _B}M_i^T \sqrt{\rho _B}}{\Tr \sqrt{\rho _B}M_i^T \sqrt{\rho _B}} 
\ket{\xi _j^i} = \frac{1}{{\rm rank} M_i}. 
\end{equation}
Note that 
$\omega _B $ is the Bob's state 
after the Alice's first measurement in the case where the given state is $\ket{\Psi}$, 
and thus, Bob only needs to detect the subspace ${\rm Ran} \omega _B$ in this case.
That is, Bob's POVM consists of $\{ \ket{\xi _j^i} \bra{\xi _j^i} \}_{j=1}^{{\rm rank} M_i}$ and 
$I_B - \sum _{j=1}^{{\rm rank } M_i} \ket{\xi _j^i} \bra{\xi _j^i}$; 
if Bob derives the measurement result corresponding to $I_B - \sum _{j=1}^{{\rm rank } M_i} \ket{\xi _j^i} \bra{\xi _j^i}$,
then, he judges that the given state is $\rho _{\rm mix}$.
We also should note that 
due to this Bob's mutually unbiased measurement,
our three-step protocol can not be reduced to a two-step one-way LOCC protocol.
If all Bob's POVMs are commutative with the eigen basis of $\omega _B$,
the whole protocol can be reduced to a one-way LOCC protocol; however,
$\omega _B$ never commutes the projection onto the mutually unbiased basis of the eigen basis of $\omega _B$.
Finally, under the above assumptions, we can write down the trace of the whole POVM element $\Tr T$ as follows,
\begin{eqnarray}
\fl \Tr T &=& \Tr \left ( \sum _{i=1}^d \sum _{j=1}^{{\rm rank} M_i}  \left ( \sqrt{M_i } \{ \sigma _A^{ij} > 0\} \sqrt{M_i } \right ) 
\otimes N_j^i \right ) \nonumber \\
\fl &=&  \sum _{i=1}^d \sum _{j=1}^{{\rm rank} M_i} \Tr \left (  \left ( \sqrt{M_i } 
\left ( \frac{\sqrt{M_i}\sqrt{\rho _A}(\ket{\xi _j^i}\bra{\xi _j^i})^T \sqrt{\rho _A}\sqrt{M_i}}{\bra{\xi_j^i}\sqrt{\rho _A}M_i^T \sqrt{\rho _A}\ket{\xi _j^i}} 
\right ) \sqrt{M_i } \right ) \otimes \ket{\xi _j^i}\bra{\xi _j^i} \right ) \nonumber \\
\fl &=& \sum _{i=1}^d \sum _{j=1}^{{\rm rank} M_i} \frac{\bra{\xi _j^i}\sqrt{\rho _A} (M_i^T)^2 \sqrt{\rho _A} 
\ket{\xi _j^i}}{\bra{\xi _j^i}\sqrt{\rho _A} M_i^T \sqrt{\rho _A} \ket{\xi _j^i}} \nonumber \\
\fl &=& \sum _{i=1}^d {\rm rank } M_i \frac{\Tr \sqrt{\rho _A}(M_i^T)^2\sqrt{\rho _A }}{\Tr \sqrt{\rho _A}M_i^T\sqrt{\rho _A}} \nonumber \\
\fl &=& \sum _{i=1}^d i \cdot \frac{\sum _{k=1}^i \lambda _k \delta _{ki}^2}{\sum _{k=1}^i \lambda _i \delta _{ki}}. \nonumber
\end{eqnarray}
In the second line of the above equalities, we used Eq.(\ref{definition of sigma _A}) (the definition of $\sigma _A^{ij}$ ) 
and the equality $\left \{  \frac{\sqrt{M_i}\sqrt{\rho _A}(\ket{\xi _j^i}\bra{\xi _j^i})^T \sqrt{\rho _A}\sqrt{M_i}}{\bra{\xi_j^i}\sqrt{\rho _A}M_i^T 
\sqrt{\rho _A}\ket{\xi _j^i}} 
> 0 \right \} = \frac{\sqrt{M_i}\sqrt{\rho _A}(\ket{\xi _j^i}\bra{\xi _j^i})^T 
\sqrt{\rho _A}\sqrt{M_i}}{\bra{\xi_j^i}\sqrt{\rho _A}M_i^T \sqrt{\rho _A}\ket{\xi _j^i}} $.
In the fourth line of the above equalities, we used the relation $\rho _A = \rho _B$ and the condition of mutually unbiased basis Eq.(\ref{mutually unbiased}).
Therefore, our problem is reduced to the optimization of 
$\sum _{i=1}^d i \cdot \frac{\sum _{k=1}^i \lambda _k \delta _{ki}^2}{\sum _{k=1}^i \lambda _i \delta _{ki}}$
over $\{ \delta _{ki} \}_{ki}$ subjected to the constraints $\delta _{ki} \ge 0$ and $\sum _{i=k}^d \delta _{ki}=1$. 
In other words, we can summarized the above discussion in the form of the following lemma.
\begin{Lemma}
For a bipartite pure state $\ket{\Psi} \in \Hi _A \otimes \Hi _B$, $\beta_{\leftrightarrow}(\ket{\Psi})$ satisfies the following inequality,
\begin{eqnarray}
\fl \beta_{\leftrightarrow}(\ket{\Psi}) 
&\le &\beta_{\widetilde{\leftrightarrow}}(\ket{\Psi}) \nonumber\\
\fl &\stackrel{\rm def}{=} &
\frac{1}{d^2}\min _{\{ \delta _{ki} \}_{1 \le k \le i \le d}} \left \{ 
\sum _{i=1}^d i \cdot \frac{\sum _{k=1}^i \lambda _k \delta _{ki}^2}{\sum _{k=1}^i \lambda _k \delta _{ki}} \ \Big| \forall k, \forall i, 
\delta _{ki} \ge 0, \ {\rm and}\ \forall k,  \sum _{i=k}^d \delta_{ki} =1 \right \},
\label{upperbound of two-way t}
\end{eqnarray}
where $\{ \lambda _k \}_{k=1}^d $ is the Schmidt coefficients of $\ket{\Psi }$, and satisfies $\lambda _k \ge \lambda _{k+1}$ for all $k$,
and the indices $(k,i)$ moves among all of the triangle region $1 \le k \le i \le d$.
\end{Lemma}

Then, the above inequality can write as $\beta_{\leftrightarrow}(\ket{\Psi}) \le \beta_{\widetilde{\leftrightarrow}}(\ket{\Psi})$.
Together with the results of the past sections,
we derive the following inequalities related to the minimum type 2 error probability for a bipartite pure state:
\begin{eqnarray}
\fl \beta_g(\ket{\Psi})=\frac{1}{d^2} \le \beta_{\rm sep}(\ket{\Psi}) && = \frac{1}{d^2}
\Bigl(\sum _{i} \sqrt{\lambda _i} \Bigr)^2
\nonumber \\
&& \le \beta_{\leftrightarrow} (\ket{\Psi}) \le \beta_{\widetilde{\leftrightarrow}}(\ket{\Psi})
\le \frac{1}{d^2}{\rm rank}\Tr _B(\ket{\Psi}\bra{\Psi}) = \beta_{\rightarrow}(\ket{\Psi}). \label{all inequalities}
\end{eqnarray}

For a two-qubit system, we can analytically calculate the exact value of
the upper-bound $\beta_{\widetilde{\leftrightarrow}}(\ket{\Psi})$, and can derive the following lemma.
\begin{Lemma}
In a two-qubit system,
\begin{equation} \label{two-way bound}
\beta_{\widetilde{\leftrightarrow} } (\ket{\Psi } ) = \frac{1}{2} - \frac{(1-\sqrt{2
\lambda })^2}{4 \left (1- \lambda \right )  },
\end{equation}
where $\{ 1-\lambda, \lambda \} $ is the Schmidt coefficient of $\ket{\Psi }$ satisfying $1 \le \lambda \le \frac{1}{2}$.
\end{Lemma}
\begin{Proof}
Without losing generality, we can write a bipartite state as $\ket{\Psi } = \sqrt{1-\lambda} \ket{00} + \sqrt{\lambda} \ket{11}$.
Then, by a straightforward calculation, we derive 
\begin{equation}
\beta_{\widetilde{\leftrightarrow}}(\ket{\Psi }) = \frac{1}{4}\min_{0 \le \delta \le 1} 
\left \{ 2 -\frac{\delta \left \{ (1-2\lambda) + \delta (1-\lambda) \right \} }{1-\delta (1-\lambda )} \right \},
\end{equation} 
where we substitute $\lambda _1 = 1- \lambda$, $\lambda _2 = \lambda $, $\delta _{11} = \delta $, $\delta _{12} = 1-\delta $, and $\delta _{22}=1$ 
into Eq.(\ref{upperbound of two-way t}).
Suppose $t'_{\widetilde{\leftrightarrow}}(\lambda, \delta ) 
\stackrel{\rm def}{=} 2 -\frac{\delta \left \{ (1-2\lambda) + \delta (1-\lambda) \right \} }{1-\delta (1-\lambda )}$.
Then, we can calculate the derivative of $t'_{\widetilde{\leftrightarrow}}(\lambda, \delta )$ as 
\begin{equation*}
\frac{ \partial t'_{\widetilde{\leftrightarrow}}(\lambda , \delta )}{\partial \delta } = - \{
(1-\lambda )\delta - (1-\sqrt{2\lambda}) \} \{ (1-\lambda )\delta
- (1+ \sqrt{2\lambda })\},
\end{equation*}
for fixed $0 \le \lambda \le \frac{1}{2} $.
Thus, under the condition $0 \le \delta \le 1$, $t'_{\widetilde{\leftrightarrow}}(\lambda,
\delta )$ attains its minimum when $\delta =
\frac{1-\sqrt{2\lambda}}{1-\lambda}$. Therefore, we derive
\begin{eqnarray*}
 \beta_{\widetilde{\leftrightarrow}}(\ket{\Psi})
&=& \frac{1}{4}\min _{0 \le \delta \le 1} t'_{\widetilde{\leftrightarrow}}(\lambda , \delta ) 
 = \frac{1}{2} -
\frac{(1-\sqrt{2\lambda })^2}{4\left (1-\lambda \right )}.
\end{eqnarray*}
\hfill $\square$
\end{Proof}
Therefore, for a two-qubit state $\ket{\Psi _{\lambda}} = \sqrt{1-\lambda}\ket{00}+\sqrt{\lambda}\ket{11}$,
the inequality (\ref{all inequalities}) can be reduced as follows,
\begin{eqnarray*}
\fl \beta_g(\ket{\Psi})=\frac{1}{4} \le \beta_{\rm sep} (\ket{\Psi } ) =\frac{1}{4}+ \frac{1}{2} \sqrt{\lambda (1-\lambda)}  \le \beta_{\leftrightarrow } (\ket{\Psi } )
&& \le \frac{1}{2} - \frac{(1-\sqrt{2
\lambda })^2}{4\left ( 1- \lambda \right )} \\
\fl && \le \frac{1}{2} = \beta_{\rightarrow} (\ket{\Psi } ),
\end{eqnarray*}
where the equality of the last inequality holds, if and only if the state is a product state or a maximally entangled state.
We present the graph of these bounds in \textbf{Figure.1}.
\begin{figure}[htbp]\label{graph}
\begin{center}
\includegraphics{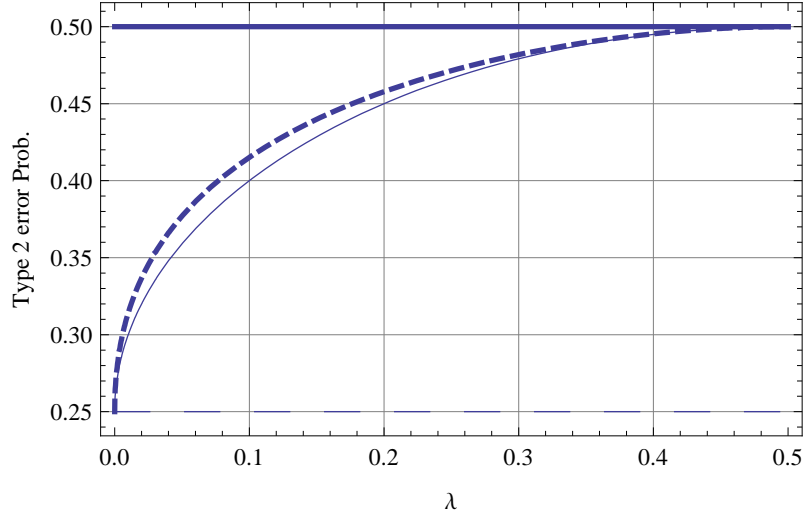}
\caption{The bound as a function of $\lambda $ (the Schmidt
coefficient of $\ket{\Psi }$ ). 
The thin line: $\beta_{\rm sep} (\ket{\Psi } )$, 
the broken line: $\beta_{\widetilde{\leftrightarrow}}(\ket{\Psi})$ 
(an upper bound of $\beta_{\leftrightarrow}(\ket{\Psi})$), 
the thick line: $\beta_{\rightarrow} (\ket{\Psi } )$,
the thin broken line: $\beta_{g} (\ket{\Psi } )$.}
\end{center}
\end{figure}
From this figure, we can see that there is a big gap between
$\beta_{\rightarrow}(\ket{\Psi})$ and
$\beta_{\leftrightarrow}(\ket{\Psi})$ and the difference between
$\beta_{\rightarrow } (\ket{\Psi }) $ and $\beta_{\rm sep}(\ket{\Psi})$ is (if
the difference exists) relatively small. Thus, for any
non-maximally entangled pure states, there is a gap between the
one-way and two-way local distinguishability at least for
two-qubit systems in terms of $\beta_{\rightarrow(\leftrightarrow
)}(\ket{\Psi})$.

In a system with a dimension of local systems $d \ge 3$, 
the optimization in the definition of $\beta_{\widetilde{\leftrightarrow}}(\ket{\Psi})$ (Eq.(\ref{upperbound of two-way t})) is too complicated to
be solved by an analytical calculation, anymore 
\footnote{In a strict sense , we can show that there exists an analytical solution for the optimization problem in Eq.(\ref{upperbound of two-way t}) 
by means of Lagrange multiplier. 
However, even for a $3 \times 3$ dimensional system, the general solution is too complicated and too ugly 
to write here. }. 
Thus, we numerically calculate the right hand side of Eq.(\ref{upperbound of two-way t})  
for a $\mathbb{C}^3 \otimes \mathbb{C}^3$ (two-qutrit) system and a $\mathbb{C}^4 \otimes \mathbb{C}^4$ system.   
For a $\mathbb{C}^3 \otimes \mathbb{C}^3$ system, we calculate Eq.(\ref{upperbound of two-way t}) for three different
one-parameter families of pure states:
\begin{enumerate}
\item $\ket{\Psi _{\lambda}} = \sqrt{1-2\lambda} \ket{11} + \sqrt{\lambda} \ket{22} + \sqrt{\lambda} \ket{33}, 
\ ( 0 \le \lambda \le \frac{1}{3})$: In this case, $\beta_g(\ket{\Psi _{\lambda}})=\frac{1}{9}$, $\beta_{\rm sep}(\ket{\Psi _{\lambda}}) = \frac{1}{9}(\sqrt{1-2\lambda} + 2\sqrt{\lambda})^2$ and 
$\beta_{\rightarrow} (\ket{\Psi _{\lambda}})=\frac{1}{3}$.
We give the results of numerical calculation of $\beta_{\widetilde{\rm sep}} (\ket{\Psi _{\lambda}} ) $ in \textbf{Figure.2}.
\begin{figure}[htbp]\label{image31}
\begin{center}
\includegraphics{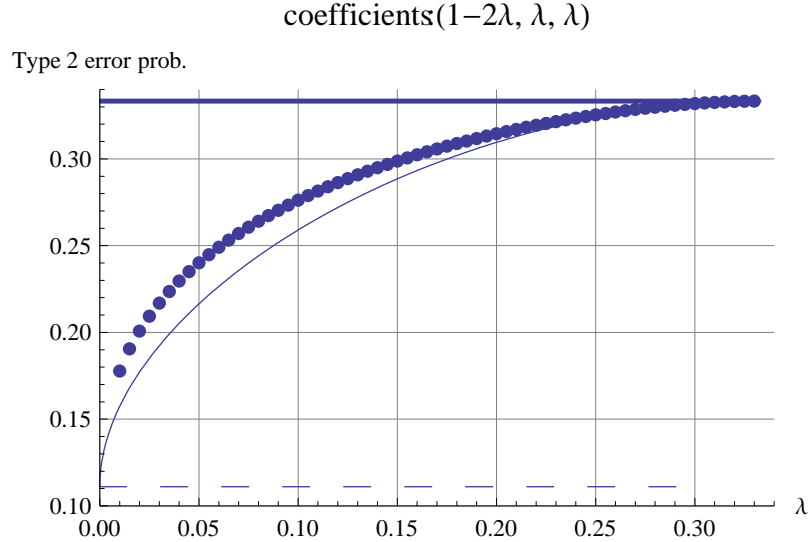}
\caption{The bound as a function of $\lambda $ for a family of states 
$\ket{\Psi _{\lambda}} = \sqrt{1-2\lambda} \ket{11} + \sqrt{\lambda} \ket{22} + \sqrt{\lambda} \ket{33}$. 
The thick broken line: results of a numerical calculation of $\beta_{\widetilde{\leftrightarrow}} (\ket{\Psi _{\lambda}} ) $ (a
lower bound of $\beta_{\leftrightarrow}(\ket{\Psi _{\lambda}})$, the thin line: 
$\beta_{\rm sep}(\ket{\Psi _{\lambda}})$, the thick line: $
\beta_{\rightarrow} (\ket{\Psi _{\lambda}} )$: the thin broken line: $\beta_g(\ket{\Psi _{\lambda}})$.}
\end{center}
\end{figure}
\item $\ket{\Psi _{\lambda}} = \sqrt{1-3\lambda} \ket{11} + \sqrt{2\lambda} \ket{22} + \sqrt{\lambda} \ket{33}, \ ( 0 \le \lambda \le \frac{1}{5})$:  
In this case, $\beta_g(\ket{\Psi _{\lambda }})=\frac{1}{9}$, $\beta_{\rm sep}(\ket{\Psi _{\lambda}}) = \frac{1}{9}(\sqrt{1-3\lambda} + (1+\sqrt{2})\sqrt{\lambda})^2$ and $\beta_{\rightarrow} (\ket{\Psi _{\lambda}} )=\frac{1}{3}$.
We give the results of a numerical calculation of $\beta_{\widetilde{\rm sep}} (\ket{\Psi _{\lambda}} ) $ in 
\textbf{Figure.3}
\begin{figure}[htbp]\label{image32}
\begin{center}
\includegraphics{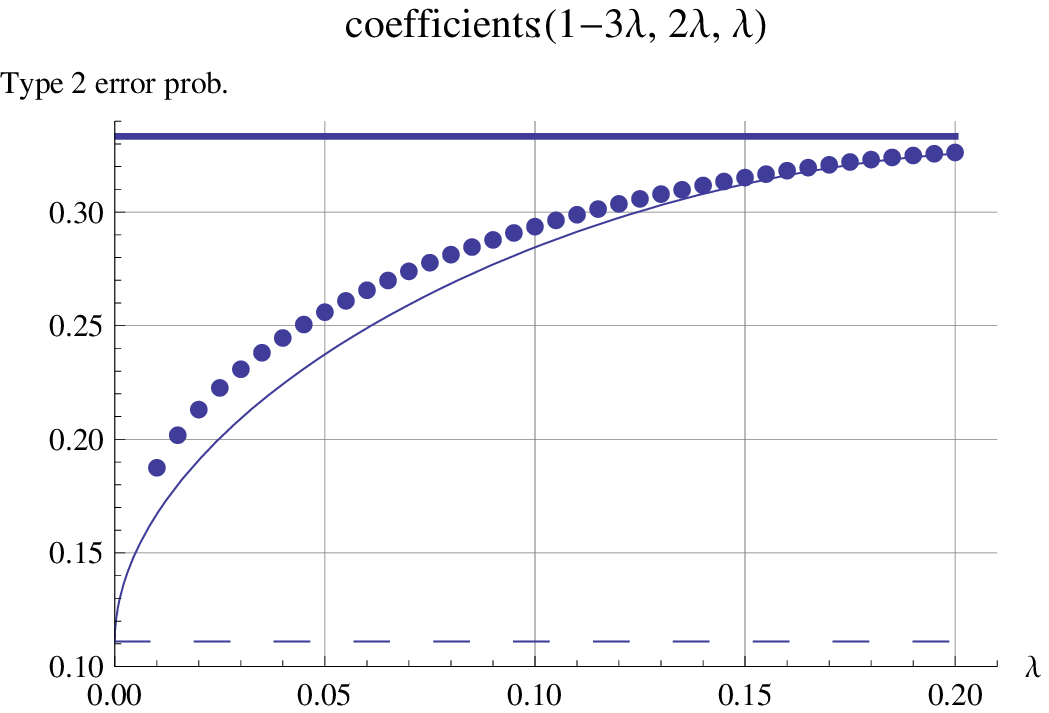}
\caption{The bound as a function of $\lambda $ for a family of states 
$\ket{\Psi _{\lambda}} = \sqrt{1-3\lambda} \ket{11} + \sqrt{2\lambda} \ket{22} + \sqrt{\lambda} \ket{33}$. 
The thick broken line: results of a numerical calculation of $\beta_{\widetilde{\leftrightarrow}} (\ket{\Psi _{\lambda}} ) $ (a
lower bound of $\beta_{\leftrightarrow}(\ket{\Psi _{\lambda}})$, the thin line: 
$\beta_{\rm sep}(\ket{\Psi _{\lambda}})$, the thick line: $
\beta_{\rightarrow} (\ket{\Psi _{\lambda}} )$, the thin broken line: $\beta_g(\ket{\Psi _{\lambda }})$.}
\end{center}
\end{figure}
\item $\ket{\Psi _{\lambda}} = \sqrt{1-4\lambda} \ket{11} + \sqrt{3\lambda} \ket{22} + \sqrt{\lambda} \ket{33}, \ ( 0 \le \lambda \le \frac{1}{7})$:  
In this case, $\beta_g(\ket{\Psi _{\lambda}})=\frac{1}{9}$, $\beta_{\rm sep}(\ket{\Psi _{\lambda}}) = \frac{1}{9}(\sqrt{1-4\lambda} + (1+\sqrt{3})\sqrt{\lambda})^2$ and 
$\beta_{\rightarrow} (\ket{\Psi _{\lambda}} )=\frac{1}{3}$.
We give the results of a numerical calculation of $\beta_{\widetilde{\rm sep}} (\ket{\Psi _{\lambda}} ) $ in 
\textbf{Figure.4}
\begin{figure}[htbp]\label{image33}
\begin{center}
\includegraphics{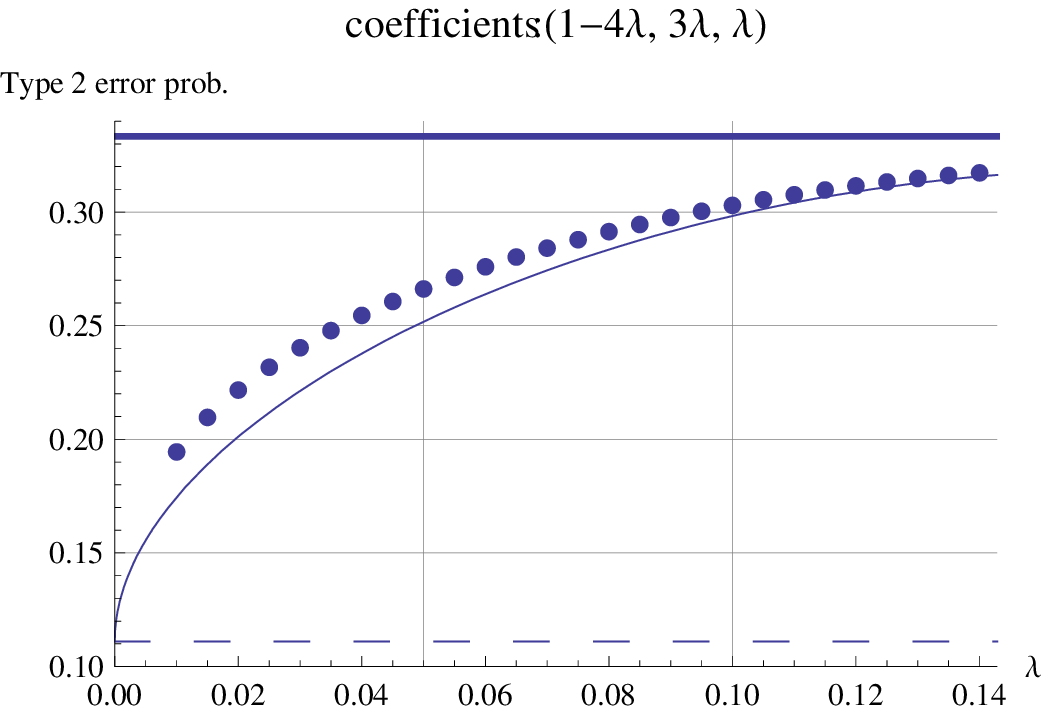}
\caption{The bound as a function of $\lambda $ for a family of states 
$\ket{\Psi _{\lambda}}=\sqrt{1-4\lambda} \ket{11} + \sqrt{3\lambda} \ket{22} + \sqrt{\lambda} \ket{33}$. 
The thick broken line: results of a numerical calculation of $\beta_{\widetilde{\leftrightarrow}} (\ket{\Psi _{\lambda}} ) $ (a
lower bound of $\beta_{\leftrightarrow}(\ket{\Psi _{\lambda}})$, the thin line: 
$\beta_{\rm sep}(\ket{\Psi _{\lambda}})$, the thick line: $
\beta_{\rightarrow} (\ket{\Psi _{\lambda}} )$, the thin broken line: $\beta_g(\ket{\Psi _{\lambda}})$.}
\end{center}
\end{figure}
\end{enumerate}
From {\bf Figures 2, 3, and 4}, we can confirm that 
the shapes of the graphs of $\beta_{\rm sep}(\ket{\Psi _{\lambda}})$, and 
$\beta_{\widetilde{\leftrightarrow}}(\ket{\Psi _{\lambda}})$  hardly depend of the choice of a one-parameter family 
$\ket{\Psi _{\lambda }}$ in $\mathbb{C}^3 \otimes \mathbb{C}^3$.
For a $\mathbb{C}^4 \otimes \mathbb{C}^4$ system, we calculate Eq.(\ref{upperbound of two-way t}) for two different
one-parameter families of pure states:
\begin{enumerate}
\item $\ket{\Psi _{\lambda}} = \sqrt{1-3\lambda} \ket{11} + \sqrt{\lambda} (\ket{22} + \ket{33}+ \ket{44}), 
\ ( 0 \le \lambda \le \frac{1}{4})$: In this case, $\beta_g(\ket{\Psi _{\lambda}})=\frac{1}{16}$, $\beta_{\rm sep}(\ket{\Psi _{\lambda}}) = \frac{1}{16}(\sqrt{1-3\lambda} + 3\sqrt{\lambda})^2$ and 
$\beta_{\rightarrow} (\ket{\Psi _{\lambda}})=\frac{1}{4}$.
We give the results of numerical calculation of $\beta_{\widetilde{\rm sep}} (\ket{\Psi _{\lambda}} ) $ in \textbf{Figure.5}.
\begin{figure}[htbp]\label{image41}
\begin{center}
\includegraphics{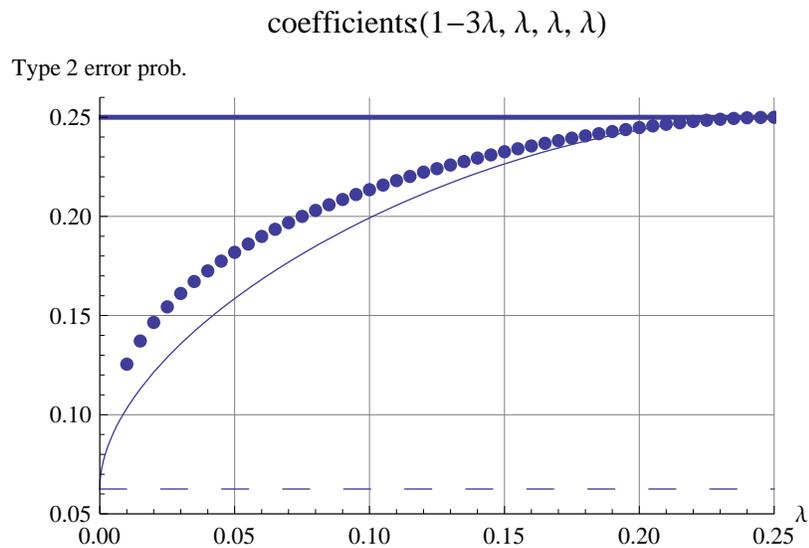}
\caption{The bound as a function of $\lambda $ for a family of states 
$\ket{\Psi _{\lambda}} = \sqrt{1-3\lambda} \ket{11} + \sqrt{\lambda} (\ket{22} + \ket{33}+ \ket{44})$. 
The thick broken line: results of a numerical calculation of $\beta_{\widetilde{\leftrightarrow}} (\ket{\Psi _{\lambda}} ) $ (a
lower bound of $\beta_{\leftrightarrow}(\ket{\Psi _{\lambda}})$, the thin line: 
$\beta_{\rm sep}(\ket{\Psi _{\lambda}})$, the thick line: $
\beta_{\rightarrow} (\ket{\Psi _{\lambda}} )$, the thin broken line: $\beta_g(\ket{\Psi _{\lambda}})$.}
\end{center}
\end{figure}
\item $\ket{\Psi _{\lambda}} = \sqrt{1-\frac{9}{2}\lambda} \ket{11} + \sqrt{2\lambda} \ket{22} + 
\sqrt{\frac{3}{2}\lambda} \ket{33} + \sqrt{\lambda}\ket{44}, \ ( 0 \le \lambda \le \frac{2}{13})$:  
In this case, $\beta_g(\ket{\Psi _{\lambda }})=\frac{1}{16}$, $\beta_{\rm sep}(\ket{\Psi _{\lambda}}) = 
\frac{1}{16} \left (\sqrt{1-\frac{9}{2}\lambda} + \left (1+\sqrt{\frac{3}{2}}+\sqrt{2} \right )\sqrt{\lambda} \right )^2$ and 
$\beta_{\rightarrow} (\ket{\Psi _{\lambda}} )=\frac{1}{4}$.
We give the results of numerical calculation of $\beta_{\widetilde{\rm sep}} (\ket{\Psi _{\lambda}} ) $ in 
\textbf{Figure.6}
\begin{figure}[htbp]\label{image42}
\begin{center}
\includegraphics{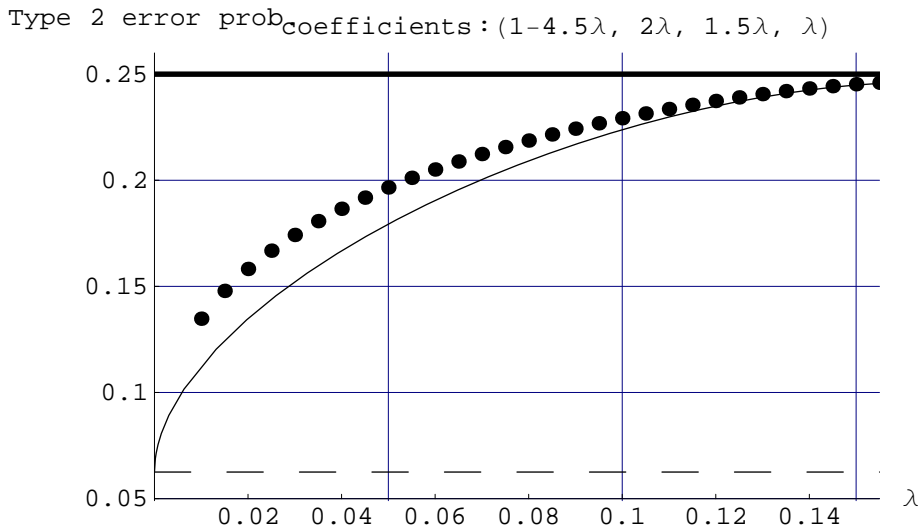}
\caption{The bound as a function of $\lambda $ for a family of states 
$\ket{\Psi _{\lambda}} = \ket{\Psi _{\lambda}} = \sqrt{1-\frac{9}{2}\lambda} \ket{11} + \sqrt{2\lambda} \ket{22} + 
\sqrt{\frac{3}{2}\lambda} \ket{33} + \sqrt{\lambda}\ket{44}$. 
The thick broken line: results of a numerical calculation of $\beta_{\widetilde{\leftrightarrow}} (\ket{\Psi _{\lambda}} ) $ (a
lower bound of $\beta_{\leftrightarrow}(\ket{\Psi _{\lambda}})$, the thin line: 
$\beta_{\rm sep}(\ket{\Psi _{\lambda}})$, the thick line: $
\beta_{\rightarrow} (\ket{\Psi _{\lambda}} )$, the thin broken line: $\beta _g (\ket{\Psi _{\lambda }})$.}
\end{center}
\end{figure}
\end{enumerate}
From {\bf Figures. 5, and 6}, we can confirm that 
the shapes of the graphs of $\beta_{\rm sep}(\ket{\Psi _{\lambda}})$, and $\beta_{\widetilde{\leftrightarrow}}(\ket{\Psi _{\lambda}})$
hardly depend of the choice of a one-parameter family $\ket{\Psi _{\lambda }}$ in $\mathbb{C}^4 \otimes \mathbb{C}^4$
as well as in $\mathbb{C}^3 \otimes \mathbb{C}^3$.
Note that, for the all above families of states, 
we choose a parameter $\lambda $ so that $\ket{\Psi _{\lambda}}$ can be converted to $\ket{\Psi _{\lambda'}}$ by LOCC for all $\lambda \ge \lambda'$,
and $\ket{\Psi _0}$ is a product state; that is, in a naive sense, 
the degree of entanglement increases monotonically when $\lambda$ increases.
From \textbf{Figures 2,3,4,5,6}, as well as for a two-qubit system (\textbf{Figure.1}),
we can see that there is always a big gap between
$\beta_{\rightarrow}(\ket{\Psi})$ and
$\beta_{\leftrightarrow}(\ket{\Psi})$ and the difference between
$\beta_{\rightarrow } (\ket{\Psi }) $ and $\beta_{\rm sep}(\ket{\Psi})$ is (if
the difference exists) relatively small for $\mathbb{C}^3 \otimes \mathbb{C}^3$ and $\mathbb{C}^4 \otimes \mathbb{C}^4$ systems. 
Moreover, since the shape of graph corresponding to $\beta_{\widetilde{\leftrightarrow}}(\ket{\Psi})$ seems not to change 
depending on a dimension of a system, we may guess that, 
for any non-maximally entangled pure states (even in a high dimensional system), there is a gap between the
one-way and two-way local distinguishability in terms of $\beta_{\rightarrow(\leftrightarrow
)}(\ket{\Psi})$. 
That is, the two-way classical communication
remarkably improves the local distinguishability compared to the
local discrimination by the one-way classical communication at least for bipartite pure states.

\section{Conclusion}
In this paper, in order to clarify the difference of the two-way LOCC and the one-way LOCC on local discrimination problems, 
we concentrated ourselves on the local discrimination
of a given bipartite state from the completely mixed state $\rho_{\rm mix}$
under the condition where the given state should be detected perfectly
while 
the previous researches \cite{VSPM01, two-state-LD} treated 
the same problem between two bipartite pure states.
We defined $\beta_{\rightarrow}(\rho)$, $\beta_{\leftrightarrow}(\rho)$, 
and $\beta_{\rm sep}(\rho)$ as the minimum error probability 
to detect the completely
mixed state by the one-way LOCC, the two-way LOCC, and the separable operation, 
respectively, under the condition that a given state $\rho$ is detected perfectly.
First, in Section \ref{separable section chap.3}, for separable operations, we showed that the minimum error probability $t_{\rm sep} (\rho)$ 
coincides with an entanglement measure called the global robustness of entanglement for a bipartite pure state except an unimportant constant term. 
Then, in Section \ref{bipartite}, 
for one-way LOCC, we showed that the minimum error probability 
$\beta_{\rightarrow}(\rho)$ coincides with the Schmidt rank for a bipartite pure state except an unimportant constant term. 
Finally, in Section \ref{two-way}, by
constructing a concrete three-step two-way LOCC discrimination protocol, we derived
an upper bound for the minimum error probability $\beta_{\leftrightarrow}(\rho)$ for a bipartite pure state. By calculating this upper bound analytically and also
numerically, we showed that $\beta_{\leftrightarrow}(\rho)$ is strictly smaller than $\beta_{\rightarrow}(\rho)$, and moreover, $\beta_{\leftrightarrow}(\rho)$
and $\beta_{\rm sep}(\rho)$ give almost the same value for a lower dimensional bipartite pure state; this
results can be seen in Figures 2,3,4,5,6. 
As a result, although there is no difference between the one-way LOCC and the two-way LOCC concerning 
local discrimination between two bipartite pure states \cite{VSPM01, two-state-LD},
we conclude that the two-way
classical communication remarkably improves the local distinguishability in comparison
with the one-way classical communication for a low-dimensional pure state at least in
the present problem setting.
Due to our quantitative comparison,
from the continuity of the second kinds of error probabilities,
a similar result should holds when 
the second state $\tilde{\rho}$ belongs to the neighborhood of the completely mixed state.
Further, we are preparing a forthcoming manuscript concerning this kind of problem in the case of multi-partite 
case in the near future \cite{future}.

\section*{Acknowledgments}
MO would like to thank Prof. Mio Murao, Dr. Damian Markham, and Dr. Shashank Virmani for helpful comments and discussions. 
MO is grateful to Professor Hiroshi Imai for his support and encouragement
The authors also appreciate Reviewr's comments.

This work was supported by Special Coordination
Funds for Promoting Science and Technology and a MEXT Grant-in-Aid for
Scientific Research on Priority,
Deepening and Expansion of Statistical Mechanical Informatics (DEX-SMI),
No. 18079014.

\appendix
\section{Proof of Eq.(\ref{average})}\label{appendix}
Now, we prove Eq.(\ref{average}), which is used in proof of Theorem \ref{theorem separable main}.
Suppose $P \stackrel{\rm def}{=} \frac{1}{2}\sum _{i\neq j}
\ket{\overline{a_{ij}}}\bra{\overline{a_{ij}}} 
\otimes \ket{\overline{b_{ij}}}\bra{\overline{b_{ij}}}$ and $Q \stackrel{\rm def}{=} \sum _{i\neq j}\{ \sum _{k \neq i,j} \lambda _k + (\sqrt{\lambda _i} - \sqrt{\lambda _j})^2\}\ket{e_if_j} \bra{e_if_j}$; that is, $\overline{T_0} =  P + Q$.
Then, by applying twirling operation over $U_{\overrightarrow{\theta}}$, we drive the following equality:
\begin{eqnarray*}
\fl &\quad & \int  _0^{2 \pi} \cdots \int _0^{2
\pi} U_{\overrightarrow{\theta}} P U_{\overrightarrow{\theta}}^{\dagger} d \theta _1 \cdots d \theta _d \nonumber  \\
\fl &=& 
(\sum _{j'=1}^d \ket{e_{j'}}\bra{e_{j'}} \otimes \ket{f_{j'}}\bra{f_{j'}} ) 
P
(\sum _{j'=1}^d \ket{e_{j'}}\bra{e_{j'}} \otimes \ket{f_{j'}}\bra{f_{j'}} ) 
\\
\fl  &\quad & +\sum _{i' \neq k'} 
\left( \ket{e_{i'}}\bra{e_{i'}}\otimes \ket{f_{k'}}\bra{f_{k'}} \right ) 
P 
\left( \ket{e_{i'}}\bra{e_{i'}}\otimes \ket{f_{k'}}\bra{f_{k'}} \right ) 
\end{eqnarray*}
This equality can be proven as follows: The action of a twirling operation (group-averaging) over a unitary representation of a compact topological group is
equal to the action of the projection onto the subspace of all invariant elements under the group action \cite{HMMOS07}.
For the action of $U_{\overrightarrow{\theta}}$ and $U_{\overrightarrow{\theta}}^{\dagger}$, 
the subspace (of operator-space $\B (\Hi)$) 
consisting of all the invariant element 
 is spanned by the operators $\{ \ket{e_jf_k}\bra{e_jf_k} \}_{j \neq k}$ and $\{ \ket{e_jf_j}\bra{e_kf_k} \}_{ij}$.
Therefore, we can easily see the above equation.
For $i \neq j$, $i'\neq k'$,
we have
\begin{eqnarray*}
\fl (\sum _{j'=1}^d \ket{e_{j'}}\bra{e_{j'}} \otimes \ket{f_{j'}}\bra{f_{j'}} ) 
\ket{\overline{a_{ij}}} \ket{\overline{b_{ij}}}
&=& \sqrt{\lambda_j}\ket{e_i f_i} -\sqrt{\lambda_i}\ket{e_j f_j} \\
\fl \left( \ket{e_{i'}}\bra{e_{i'}}\otimes \ket{f_{k'}}\bra{f_{k'}} \right ) 
\ket{\overline{a_{ij}}} \ket{\overline{b_{ij}}}
&=&
\delta_{i',i}\delta_{k',j}
(\lambda_j\lambda_j)^{\frac{1}{4}}\ket{e_i f_j} 
-
\delta_{i',j}\delta_{k',i}
(\lambda_j\lambda_j)^{\frac{1}{4}}\ket{e_j f_i} .
\end{eqnarray*}
Since
\begin{eqnarray*}
\fl &&(\sqrt{\lambda_j}\ket{e_i f_i} -\sqrt{\lambda_i}\ket{e_j f_j})
(\sqrt{\lambda_j}\bra{e_i f_i} -\sqrt{\lambda_i}\bra{e_j f_j})\\
\fl &=&
\lambda_j\ket{e_i f_i}\bra{e_i f_i}
+\lambda_i\ket{e_j f_j}\bra{e_j f_j}
- \sqrt{\lambda_j}\sqrt{\lambda_i}
\ket{e_i f_i}\bra{e_j f_j}
- \sqrt{\lambda_i}\sqrt{\lambda_j}
\ket{e_j f_j}\bra{e_i f_i},
\end{eqnarray*}
we obtain
\begin{eqnarray*}
\fl &\quad & \int  _0^{2 \pi} \cdots \int _0^{2
\pi} U_{\overrightarrow{\theta}} P U_{\overrightarrow{\theta}}^{\dagger} d \theta _1 \cdots d \theta _d \nonumber  \\
\fl &=& 
\sum_{i \neq j}
\Bigl(\lambda_j\ket{e_i f_i}\bra{e_i f_i}
+\lambda_i\ket{e_j f_j}\bra{e_j f_j}
- \sqrt{\lambda_j}\sqrt{\lambda_i}
\ket{e_i f_i}\bra{e_j f_j}
- \sqrt{\lambda_i}\sqrt{\lambda_j}
\ket{e_j f_j}\bra{e_i f_i}\Bigr)\\
\fl &&
+ \sum _{i\neq j} \sqrt{\lambda _i\lambda _j}\ket{e_if_j}\bra{e_if_j} 
\\
\fl &=& (\sum _i^d \ket{e_if_i}\bra{e_if_i}) - (\sum _{i=1}^d \sqrt{\lambda _i} \ket{e_if_i})(\sum _{i=1}^d \sqrt{\lambda _i} \bra{e_if_i})
+ \sum _{i\neq j} \sqrt{\lambda _i\lambda _j}\ket{e_if_j}\bra{e_if_j} .
\end{eqnarray*}
In the same way, we can also show the equality $\int  _0^{2 \pi} \cdots \int _0^{2
\pi} U_{\overrightarrow{\theta}} Q U_{\overrightarrow{\theta}}^{\dagger} d \theta _1 \cdots d \theta _d = Q$; $Q$ is invariant
under the twirling operation.

Finally, we can calculate $\int  _0^{2 \pi} \cdots \int _0^{2
\pi} U_{\overrightarrow{\theta}} \overline{T_0} U_{\overrightarrow{\theta}}^{\dagger} d \theta _1 \cdots d \theta _d$ as follows:
\begin{eqnarray*}
\fl &\quad & \int  _0^{2 \pi} \cdots \int _0^{2
\pi} U_{\overrightarrow{\theta}} \overline{T_0} U_{\overrightarrow{\theta}}^{\dagger} d \theta _1 \cdots d \theta _d 
= \int  _0^{2 \pi} \cdots \int _0^{2
\pi} U_{\overrightarrow{\theta}} (P+Q) U_{\overrightarrow{\theta}}^{\dagger} d \theta _1 \cdots d \theta _d \\
\fl &=& 
(\sum _i^d \ket{e_if_i}\bra{e_if_i}) - (\sum _{i=1}^d \sqrt{\lambda _i} \ket{e_if_i})(\sum _{i=1}^d \sqrt{\lambda _i} \bra{e_if_i})
+ \sum _{i\neq j} \sqrt{\lambda _i\lambda _j}\ket{e_if_j}\bra{e_if_j} \\
\fl &\quad & + \sum _{i\neq j}\{ \sum _{k \neq i,j} \lambda _k + (\sqrt{\lambda _i} - \sqrt{\lambda _j})^2\}\ket{e_if_j} \bra{e_if_j} \\
\fl &=& (\sum _i^d \ket{e_if_i}\bra{e_if_i}) + (\sum _{i \neq j} \ket{e_if_j}\bra{e_if_j})  - (\sum _{i=1}^d \sqrt{\lambda _i} \ket{e_if_i})(\sum _{i=1}^d \sqrt{\lambda _i} \bra{e_if_i}) \\
\fl &\quad & 
- \sum _{i\neq j} \sqrt{\lambda _i\lambda _j}\ket{e_if_j}\bra{e_if_j} \\
\fl &=& I - T,
\end{eqnarray*} 
which proves Eq.(\ref{average}). \hfill $\square$
%%%%%%%%%%%%%%%%%%%%%%%%%%%%%%%%%%%%%%%%%%%%%%%%%%%%%%%%%%%%%%%%%%%%%%%%%%%%%%

\end{document}